\documentclass[twocolumn,showpacs,preprintnumbers,superscriptaddress,nofootinbib]{revtex4-1}
\usepackage{graphicx,fancybox}
\usepackage{amsmath,amssymb,bm}

\usepackage[T1]{fontenc}
\usepackage[table]{xcolor}
\usepackage{ulem}

\begin{document}

\title{Distribution of Energy-Momentum Tensor around a Static Quark \\
in the Deconfined Phase of SU(3) Yang-Mills Theory  }

\author{Ryosuke Yanagihara}
\email{yanagihara@kern.phys.sci.osaka-u.ac.jp}
\affiliation{Department of Physics, Osaka University,
Toyonaka, Osaka 560-0043, Japan}

\author{Masakiyo Kitazawa}
\email{kitazawa@phys.sci.osaka-u.ac.jp}
\affiliation{Department of Physics, Osaka University,
Toyonaka, Osaka 560-0043, Japan}
\affiliation{J-PARC Branch, KEK Theory Center,
Institute of Particle and Nuclear Studies, KEK,
203-1, Shirakata, Tokai, Ibaraki, 319-1106, Japan }
  
\author{Masayuki Asakawa}
\email{yuki@phys.sci.osaka-u.ac.jp}
\affiliation{Department of Physics, Osaka University,
Toyonaka, Osaka 560-0043, Japan}

\author{Tetsuo Hatsuda}
\email{thatsuda@riken.jp}
\affiliation{RIKEN Interdisciplinary Theoretical and 
Mathematical Sciences Program (iTHEMS), RIKEN, 
Wako 351-0198, Japan}

\begin{abstract}
 Energy momentum tensor (EMT) characterizes the response of
 the vacuum as well as the thermal medium
 under the color electromagnetic fields.
 We define the EMT by means of the gradient flow formalism
 and study its spatial distribution 
 around a static quark in the deconfined phase of 
 SU(3) Yang-Mills theory on the lattice.
 Although  no significant  difference 
 can be seen between the EMT distributions 
 in the radial and transverse directions
 except for the sign, the temporal component 
 is substantially different from the 
 spatial ones near the critical temperature $T_c$. 
 This is in contrast to the 
 prediction of the leading-order thermal perturbation theory. 
 The lattice data of the EMT distribution also indicate 
 the thermal screening 
 at long distance and the perturbative behavior at short distance.
\end{abstract}

\preprint{J-PARC-TH-0229, RIKEN-iTHEMS-Report-20}

\maketitle

\section{Introduction} 
\label{sec:introduction}

To study  complex quantum systems  such as the Yang-Mills (YM) theory,
it is customary to introduce test probe(s) and analyze the response.
The Wilson loop is one of such probes 
whose measurement in 
YM theory
provides information on the static quark--anti-quark system
that is closely related to 
the confinement property in YM vacuum \cite{Bali:2000gf}.
Thanks to the recent development 
of the gradient-flow 
method~\cite{Narayanan:2006rf,Luscher:2010iy,Luscher:2011bx} 
and its application to the energy-momentum tensor (EMT) 
$\mathcal{T}_{\mu\nu}(x)$~\cite{Suzuki:2013gza,Makino:2014taa,Hieda:2016xpq,Harlander:2018zpi}, 
it became possible to study the
gauge-invariant structure of the  flux tube 
between the quark and anti-quark in the confining phase  
through the spatial distribution of EMT 
under the Wilson loop~\cite{Yanagihara:2018qqg,Yanagihara:2019foh}.

The purpose of the present  paper  is to extend  the above idea and 
to explore the EMT distribution around a static quark in YM theory.
As a first step,
we consider the deconfined phase 
above the critical temperature $T_c$ of the SU(3) YM theory in the range
of temperature $1.2 \le T/T_c \le 2.6$ and measure the EMT distribution
around the Polyakov loop.
The EMT with the gradient flow has been  
used to study thermodynamics of 
YM theory~\cite{Asakawa:2013laa,Kitazawa:2016dsl,Kitazawa:2017qab,Iritani:2018idk,Hirakida:2018uoy,Kitazawa:2019otp} 
and of QCD~\cite{Taniguchi:2016ofw,Taniguchi:2020mgg}.  
However, the observables in these studies are 
limited to global quantities such as 
the pressure, energy density, entropy density, and the specific heat.
On the other hand, we 
focus on the local observable 
in this study and examine the following questions:
(i)  How are the energy density and the stress tensor distributed 
around the static quark?, (ii) How are the distributions modified 
as a function of temperature?, 
and (iii) How can one extract parameters 
such as the running coupling and the Debye screening mass
from the distributions?  

The organization of the present paper is as follows.
In Sec.~\ref{sec:emt_general}, we briefly review 
the definition of EMT and its property in the 
spherical coordinate system.
In Sec.~\ref{sec:gradient_emt}, we introduce the EMT operator 
on the lattice and its correlation with the Polyakov loop operator.
In Sec.~\ref{sec:setup}, 
we discuss the numerical procedure 
and lattice setup to analyze the EMT operator around a static quark
on the lattice.
Numerical results  and their physical implications 
are given in Sec.~\ref{sec:result}.
Sec.~\ref{sec:summary} is devoted to the summary and conclusion.
In Appendix A, we discuss the procedure to make the tree-level improvement 
of the correlation between the EMT and the Polyakov loop on the lattice.
In  Appendix B, the leading-order perturbative analysis 
of the correlation is presented 
using the high temperature effective field theory.

\section{EMT around a static charge}
\label{sec:emt_general} 

The stress tensor $\sigma_{ij}\,(i,j = 1,2,3)$ is related to the
spatial component of EMT, $\mathcal{T}_{ij}$, as~\cite{Landau}
\begin{align}
 \sigma_{ij} = - \mathcal{T}_{ij}.
\end{align}
The force per unit area ${\cal F}_i$
acting on a surface with the normal vector $n_i$ 
is given by the stress tensor as
\begin{eqnarray}
 {\cal F}_i =  \sigma_{ij}n_j =  - {\cal T}_{ij} n_j.
  \label{eq:F=Tn}
\end{eqnarray} 
The local principal axes $n_j^{(k)}$ and the corresponding eigenvalues
$\lambda_k$
of the local stress tensor are obtained
by solving the eigenvalue problem:
\begin{align}
 {\cal T}_{ij}n_j^{(k)}=\lambda_k n_i^{(k)} \quad (k=1,2,3).
 \label{eq:EV}
\end{align}
The strength of the force per unit area
along $n_i^{(k)}$ is given by the absolute values 
of the eigenvalue $\lambda_k $.
Neighboring volume elements separated 
by a surface with the normal
vector $n_i^{(k)}$ pull (push)  each other
for $\lambda_k <0 $ ($\lambda_k >0 $) across the wall.
Note that the three principal axes $n_i^{(k)}$
are orthogonal with one another because $\sigma_{ij}$ is a symmetric tensor.

\begin{figure}
 \centering
 \includegraphics[width=0.40\textwidth, clip]{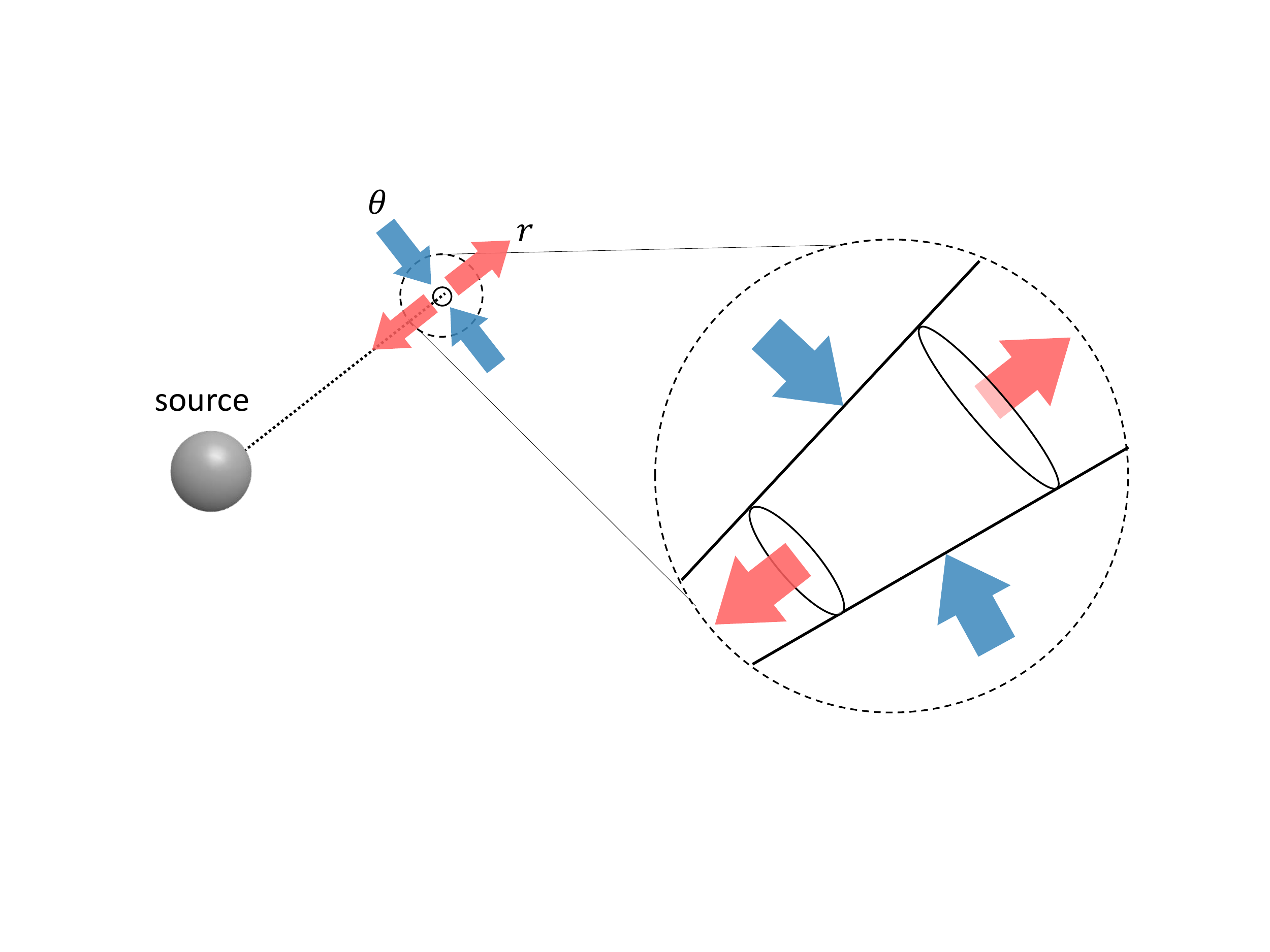}
 \caption{
 Stress acting on infinitely small volume element 
 under the existence of a single static charge (source).
 In the case of the classical electromagnetism, a small volume element 
 at a distance of $r$ from the charge is pulled 
 along the radial direction from 
 the neighboring volume elements 
 while it is pushed in the transverse direction.
 }
 \label{fig:stress_pic}
\end{figure}

For a system with a single static source,
it is convenient to use the spherical 
coordinate system $(r,\theta, \varphi)$
with the radial coordinate $r=|\bm{x}|$, and 
the polar and azimuthal angles $\theta$ and $\varphi$.
The spherical symmetry allows us to diagonalize 
the static EMT in Euclidean spacetime  $\mathcal{T}_{\mu \nu}(\bm{x})$
in this coordinate system as
\begin{align}
 \mathcal{T}_{\gamma\gamma'}(\bm{x})&=
 \mathrm{diag}(\mathcal{T}_{44}(r), \mathcal{T}_{rr}(r), 
 \mathcal{T}_{{\theta\theta}}(r))
 \label{eq:stress_separate}
\end{align}
where $\gamma,\gamma'={4,r,\theta}$.
Due to the spherical symmetry, 
the azimuthal component degenerates with the polar component,  
$\mathcal{T}_{\varphi\varphi}(r)= \mathcal{T}_{\theta\theta}(r)$,
so that only independent components are given 
in Eq.~(\ref{eq:stress_separate}).

In the Abelian case,  
EMT is given by the Maxwell stress-energy tensor~\cite{Landau},
$ \mathcal{T}^\mathrm{Maxwell}_{\mu\nu}
=  {F}_{\mu\rho } {F}_{\nu \rho} - \frac{1}{4} \delta_{\mu\nu}
{F}_{\rho\sigma} {F}_{\rho\sigma}$
with the field strength ${F}_{\mu\nu}$.
When a static charge 
is placed  at the origin,  the EMT is denoted by
\begin{align}
\mathcal{T}_{\gamma\gamma'}^{\rm Maxwell}
=\frac{1}{2}\mathrm{diag}(-\vec{E}^2, -\vec{E}^2, \vec{E}^2),
\label{eq:Maxwell}
\end{align}
with  ${E}_i(\bm{x})$ being  the electric field.
The spatial structure of Eq.~(\ref{eq:Maxwell})
is illustrated in Fig.~\ref{fig:stress_pic} where 
the neighboring volume elements around the static electric charge 
pull (push) each other  along the radial (angular) direction.
In a static system, the force acting on a volume element through
its surface should be balanced.
This property is guaranteed by the momentum conservation
$\partial_i \mathcal{T}_{ij}=0$ together with the Gauss theorem.

\section{EMT for SU(3) Yang-Mills theory on the lattice}
\label{sec:gradient_emt}

\subsection{YM gradient flow}

We consider the pure SU(3) YM gauge theory in the four-dimensional Euclidean
space defined by the action,
\begin{align}
 S_\mathrm{YM} 
 = \frac{1}{4g_0^2}\int d^4x\,G_{\mu\nu}^a(x)G_{\mu\nu}^a(x).
 \label{eq:flow_action}
\end{align}
Here $g_0$ is a bare gauge coupling  
and $G_{\mu\nu}^a(x)$ is the field strength 
composed of the fundamental gauge field 
$ A_{\mu}^a(x) $.
The YM gradient flow evolves the gauge field along the 
fictitious fifth dimension $t$ introduced in addition to 
the ordinary four Euclidean
dimensions $x$ through 
the flow equation ~\cite{Narayanan:2006rf,Luscher:2010iy,Luscher:2011bx},
\begin{align}
 \frac{dA_\mu^a(t,x)}{dt} 
 = -g_0^2\frac{\delta S_\mathrm{YM}(t)}{\delta A_\mu^a(t,x)}.
 \label{eq:GF}
\end{align}
The flowed YM action $S_\mathrm{YM}(t)$ 
in the $(4+1)$-dimensional coordinate
is constructed by substituting the flowed 
gauge field $A_\mu^a(t,x)$ in Eq.~(\ref{eq:flow_action}) with
an initial condition, $A_\mu^a(t=0,x)=A_\mu^a(x)$.

An important feature of the gradient flow for $t>0$ is
that any composite operators composed of flowed gauge fields are
UV finite even at equal spacetime point~\cite{Luscher:2011bx,Hieda:2016xpq}.
This is a consequence of the smoothing of the gauge fields in the four
dimensional Euclidean space within the range $\sim\sqrt{2t}$.
In addition, in the small $t$ limit, composite local operators 
are represented by the local operators 
of the ordinary gauge theory at $t=0$.
These properties lead us to 
the renormalized EMT operator defined with the small $t$
expansion~\cite{Suzuki:2013gza}:
\begin{eqnarray}
 {\cal T}_{\mu\nu}^{\rm R} (x)
  &= &\lim_{t\to0} {\cal T}_{\mu\nu}(t,x) , \label{eq:t} \\ 
 {\cal T}_{\mu\nu}(t,x)  &= &
 c_1(t) U_{\mu\nu}(t,x)
 \nonumber \\
  &&+ 4c_2(t) \delta_{\mu\nu}
  \left[E(t,x)-\left\langle E(t,x)\right\rangle_0 \right] ,
 \label{eq:T}
\end{eqnarray}
where $\langle E(t,x) \rangle_0$   is the vacuum expectation value
of $E(t,x)$.
The dimension-four gauge-invariant operators 
on the right hand side of Eq.~(\ref{eq:T}) are
given by \cite{Suzuki:2013gza}
\begin{align}
  E(t,x) &= \frac{1}{4} G_{\mu\nu}^a(t,x)G_{\mu\nu}^a(t,x),
  \label{eq:E} 
  \\
  U_{\mu\nu}(t,x) &= G_{\mu\rho}^a (t,x)G_{\nu\rho}^a (t,x)
  - \delta_{\mu\nu} E(t,x), 
  \label{eq:U}
\end{align}
where $G_{\mu\nu}^a (t,x)$ is the field strength composed of 
the flowed gauge field.
Because of the vacuum subtraction in Eq.~(\ref{eq:T}), 
$\langle {\cal T}_{\mu\nu}^{\rm R}(x) \rangle_0$ vanishes.
The coefficients $c_1(t)$ and $c_2(t)$ have been calculated
perturbatively in 
Refs.~\cite{Suzuki:2013gza,Harlander:2018zpi,Iritani:2018idk} 
for small $t$.
We use two-loop perturbative
coefficients~\cite{Harlander:2018zpi,Iritani:2018idk}
for the construction of EMT throughout this study.

\subsection{EMT around a static heavy quark}
To describe a static quark $Q$ on the lattice,
we introduce the Polyakov loop  at the origin, $\Omega(\bm{0})$.
Then the expectation value of Eq.~(\ref{eq:T})
around $Q$ is given by
\begin{align}
 \langle \mathcal{T}_{\mu\nu}(t,x) \rangle_Q
 = \frac{\langle  \mathcal{T}_{\mu\nu}(t,x)\mathrm{Tr}\Omega(\bm{0})\rangle}
 {\langle \mathrm{Tr}\Omega(\bm{0}) \rangle}
 -\langle  \mathcal{T}_{\mu\nu}(t,x)\rangle.
 \label{eq:cor_pol_emt0}
\end{align}
We note that Eq.~(\ref{eq:cor_pol_emt0}) is well-defined only when
the $Z_3$ symmetry in SU(3) YM theory is spontaneously broken:
In the $Z_3$ unbroken phase, both numerator and denominator
of the first term on the right hand side vanish exactly.
This is the reason why we focus on the system in the $Z_3$ broken phase
above $T_c$ in this paper.
In practice, we choose the state with  the Polyakov loop being real
among the three equivalent $Z_3$ states in the deconfined phase.

The renormalized EMT distribution around $Q$ is obtained after 
taking the double extrapolation,
\begin{align}
 \langle \mathcal{T}^\mathrm{R}_{\mu\nu}(x) \rangle_Q
 = \lim_{t \rightarrow 0} \lim_{a \rightarrow 0}
 \langle \mathcal{T}_{\mu\nu}(t,x) \rangle_Q.
 \label{eq:w-ext}
\end{align}
In our actual analysis, we extract the renormalized EMT distribution
by fitting the lattice data
with the following functional form~\cite{Kitazawa:2016dsl,Kitazawa:2017qab}:
\begin{align}
 \langle \mathcal{T}_{\mu\nu}(t,x) \rangle_Q
 = \langle \mathcal{T}^\mathrm{R}_{\mu\nu}(x) \rangle_Q
 + b_{\mu\nu}{(t)}a^2 + c_{\mu\nu}t + d_{\mu\nu}t^2, 
 \label{eq:double_lim}
\end{align}
where the contributions from discretization effects ($b_{\mu\nu}$)  
as well as the  dimension-six and -eight 
operators ($c_{\mu\nu}$ and $d_{\mu\nu}$) 
are considered.

To perform the double extrapolation reliably,
the smearing radius  $\rho\equiv \sqrt{2t}$
needs to be larger than the 
lattice spacing to suppress the discretization error. At the same time, 
$\rho$ should be smaller than half the temporal size 
$1/2T$ with temperature $T$
as well as the distance from the source ($r$)
to avoid the overlap of operators.
Therefore we require 
\begin{align}
 a/2 \lesssim \rho
 \lesssim \mathrm{min}\Bigl(r,\frac{1}{2T}\Bigr).
 \label{eq:flow_require}
\end{align}
The lattice data to be fitted by Eq.~(\ref{eq:double_lim}) 
should be within this window.
As will be discussed in Sec.~\ref{sec:double},
we impose more stringent  conditions for the range of $t$ 
in our numerical analysis.

\section{Lattice setup}
\label{sec:setup}

\subsection{Gauge configurations}
Numerical simulations in SU(3) YM theory were performed 
on the four dimensional Euclidean lattice with the Wilson gauge action and 
the periodic boundary conditions at four different temperatures
$1.20T_c,\,1.44T_c,\,2.00T_c$, and $2.60T_c$.
The simulation parameters for each $T$ are summarized 
in Table~\ref{table:params}.
The inverse coupling $\beta=6/g_0^2$ is related to the 
lattice spacing $a$ determined by  the reference 
scale $w_0$~\cite{Kitazawa:2016dsl,Borsanyi:2012zs}.
The spatial and temporal lattice sizes, $N_s$ and $N_\tau$
 together with  the number of configurations $N_\mathrm{conf}$
are also summarized in Table~\ref{table:params}.
All lattices have the same aspect ratio $N_s/N_\tau=4$.

\begin{table}
 \centering
 \caption{Simulation parameters for the four temperatures: 
 The spatial lattice size $N_s$,
 the temporal lattice size $N_\tau$,
 $\beta=6/g_0^2$,
 the lattice spacing $a$.
 $N_{\mathrm{conf}}$ represents the number of configurations.}
 \begin{tabular}{cccccr}
  \hline \hline
  $T/T_c$ & $N_s$ & $N_\tau$ & $\beta$ & $a~[\mathrm{fm}]$ & $N_{\mathrm{conf}}$ \\
  \hline
  1.20 & 40 & 10 & 6.336 & 0.0551 & 500 \\
  & 48 & 12 & 6.467 & 0.0460 & 650 \\
  & 56 & 14 & 6.581 & 0.0394 & 840 \\
  & 64 & 16 & 6.682 & 0.0344 & 1,000 \\
  & 72 & 18 &6.771 & 0.0306 & 1,000 \\
  \hline
  1.44 & 40 & 10 & 6.465 & 0.0461 & 500 \\
  & 48 & 12 & 6.600 & 0.0384 & 650 \\
  & 56 & 14 &  6.716 & 0.0329 & 840 \\
  & 64 & 16 & 6.819 & 0.0288 & 1,000 \\
  & 72 & 18 & 6.910 & 0.0256 & 1,000 \\
  \hline
  2.00 & 40 & 10 & 6.712 & 0.0331 & 500 \\
  & 48 & 12 & 6.853 & 0.0275 & 650 \\
  & 56 & 14 & 6.973 & 0.0236 & 840 \\
  & 64 & 16 & 7.079 & 0.0207 & 1,000 \\
  & 72 & 18 & 7.173 & 0.0184 & 1,000 \\
  \hline
  2.60 & 40 & 10 & 6.914 & 0.0255 & 500 \\
  & 48 & 12 & 7.058 & 0.0212 & 650 \\
  & 56 & 14 & 7.182 & 0.0182 & 840 \\
  & 64 & 16 & 7.290 & 0.0159 & 1,000 \\
  & 72 & 18 & 7.387 & 0.0141 & 1,000 \\
  \hline \hline
 \end{tabular}
 \label{table:params}
\end{table}

The gauge configurations are generated by the pseudo-heat-bath method
followed by five over-relaxations.
Each measurement is separated by 200 sweeps.
Statistical errors are estimated by the jackknife method with 20
jackknife bins.
We employ the Wilson gauge action for $S_\mathrm{YM}(t)$
in the flow equation Eq.~(\ref{eq:GF}) and the clover type
representation for the field strength $G_{\mu\nu}(t,x)$.
The numerical solution of the gradient flow equation
is obtained by the third order Runge-Kutta method.

In order to suppress the statistical noise,
we apply the multi-hit procedure in the measurement of the Polyakov
loop by replacing every temporal 
link by its thermal average~\cite{Parisi:1983hm}.
The choice of the temporal argument $x_4$ of EMT 
in Eq.~(\ref{eq:cor_pol_emt0}) is arbitrary. Therefore,
we average EMT over the temporal direction
to reduce the statistical error.

\subsection{Discretization effect}
\label{subsec:disc}

\begin{figure}[b]
 \centering
 \includegraphics[width=0.40\textwidth, clip]{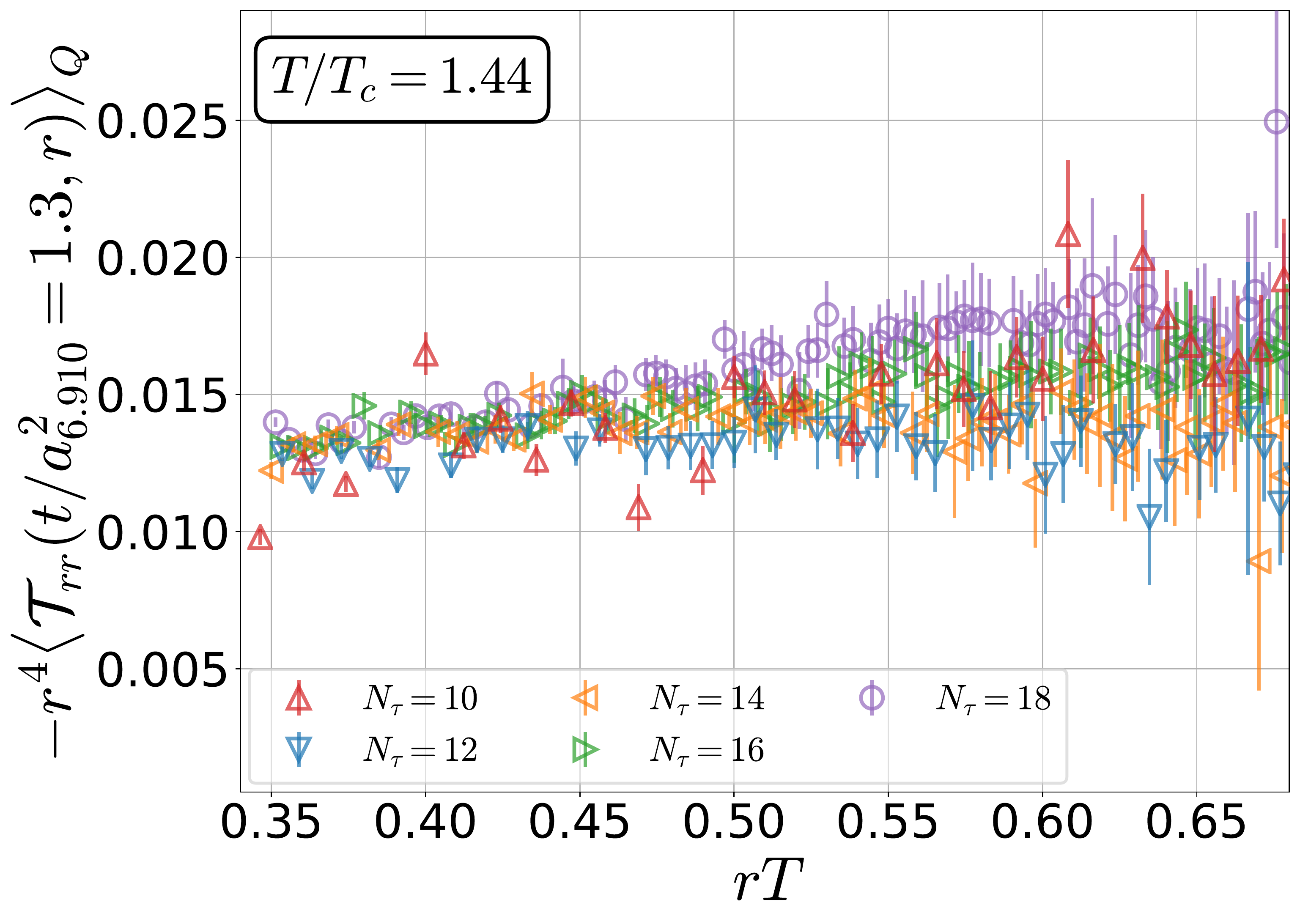}
 \caption{
 Distribution of 
 $-r^4\langle \mathcal{T}_{rr}(t/a_{6.910}^2=1.3,r) \rangle_Q$
 as functions of $rT$ at $T/T_c=1.44$.
 }
 \label{fig:finite_size_main}
\end{figure}

The EMT in the spherical coordinate system on the lattice reads
\begin{align}
 &  T \sum_{x_4} \langle \mathcal{T}_{\gamma \gamma'}(t, \bm{x}, x_4) \rangle_{Q}\notag
 \\
 &=
 \mathrm{diag} ( \langle \mathcal{T}_{44} (t,r)\rangle_{Q},
 \langle \mathcal{T}_{rr} (t,r)\rangle_{Q},
 \langle \mathcal{T}_{\theta\theta} (t,r)\rangle_{Q}).
 \label{eq:EMT-lat}
\end{align}
The behavior of the EMT distribution close to  
the source $Q$ is affected by the 
violation of rotational symmetry owing to lattice discretization.
As an example, we show in Fig.~\ref{fig:finite_size_main} the 
distribution 
of $-r^4\langle \mathcal{T}_{rr}(t/a_{6.910}^2=1.3,r)\rangle_Q$ as
a function of $rT$ at $T/T_c=1.44$, where $a_{6.910}$ is 
the lattice spacing of the finest lattice at this temperature.
The figure shows the oscillating behavior of the numerical results
 becomes more prominent on coarser lattices.
In this study, we use the lattice data only 
for $N_{\tau} \ge 12$ for the continuum extrapolation to suppress
the discretization errors.
In  Appendix~\ref{sec:append_disc},
we consider an alternative analysis that performs
the tree-level improvement of the numerical results and uses them
for the continuum extrapolation with the $N_{\tau}=10$ data.
As discussed there, we  confirm that the results 
in both cases are the same within the errors.

\subsection{Double extrapolation}
\label{sec:double}

The double extrapolation Eq.~(\ref{eq:w-ext}) consists of
two steps: (I) the continuum ($a\to0$) extrapolation, and
(II) $t\to0$ extrapolation.
In this subsection, we demonstrate
these procedures by using the lattice data at $T/T_c=1.44$ as an example.

\begin{figure}[h]
 \centering
 \includegraphics[width=0.4\textwidth, clip]{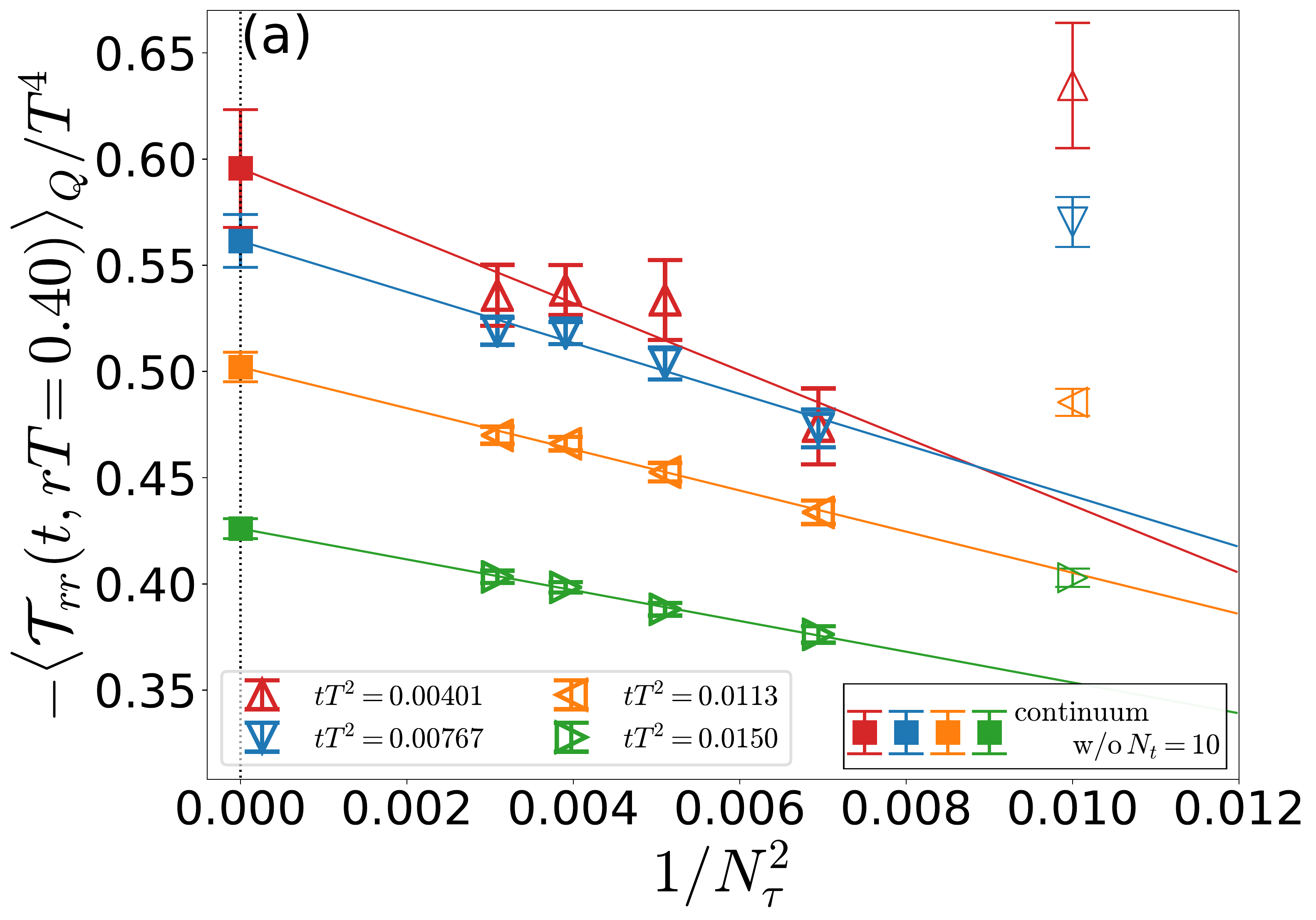}
 \includegraphics[width=0.4\textwidth, clip]{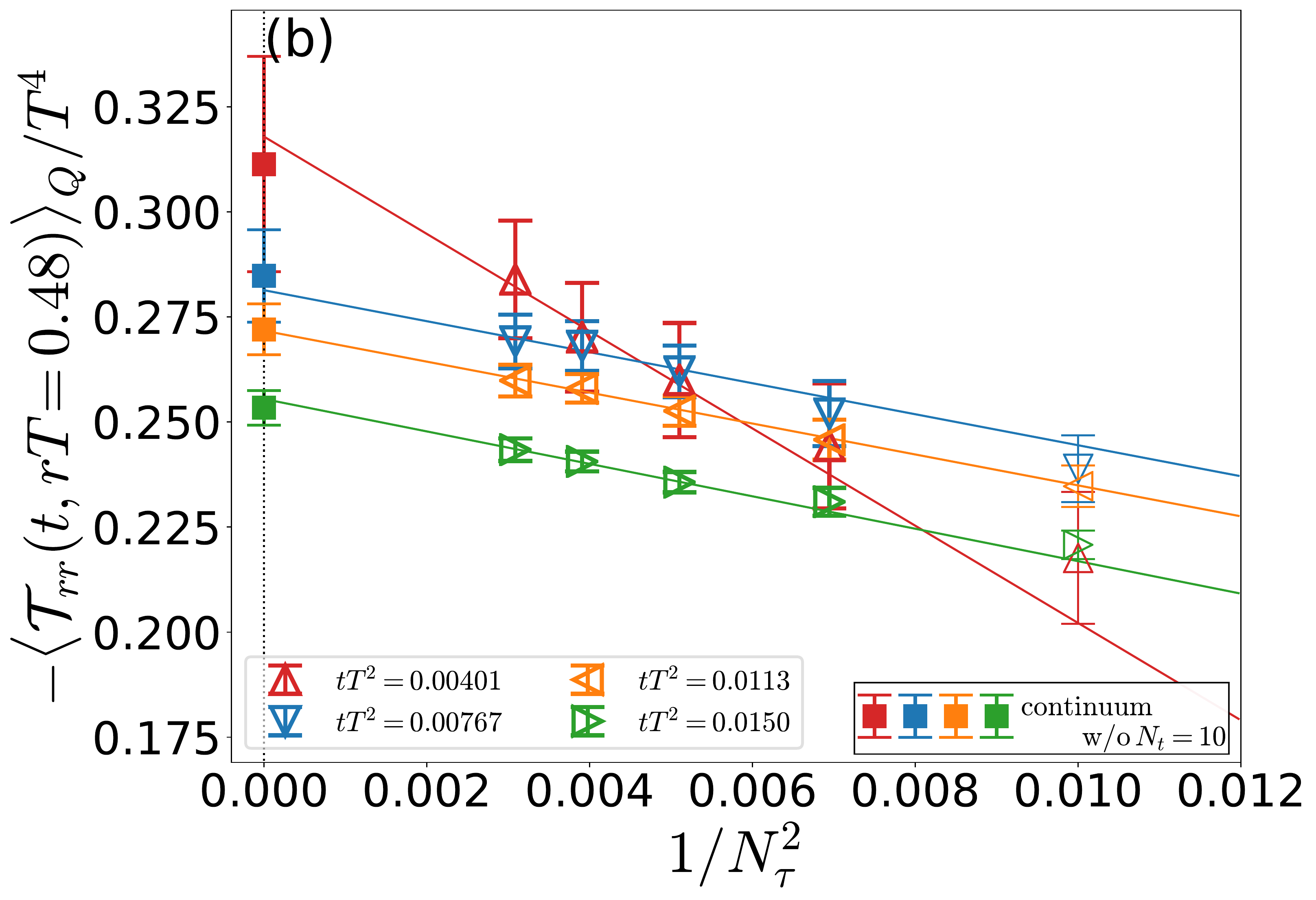}
 \includegraphics[width=0.4\textwidth, clip]{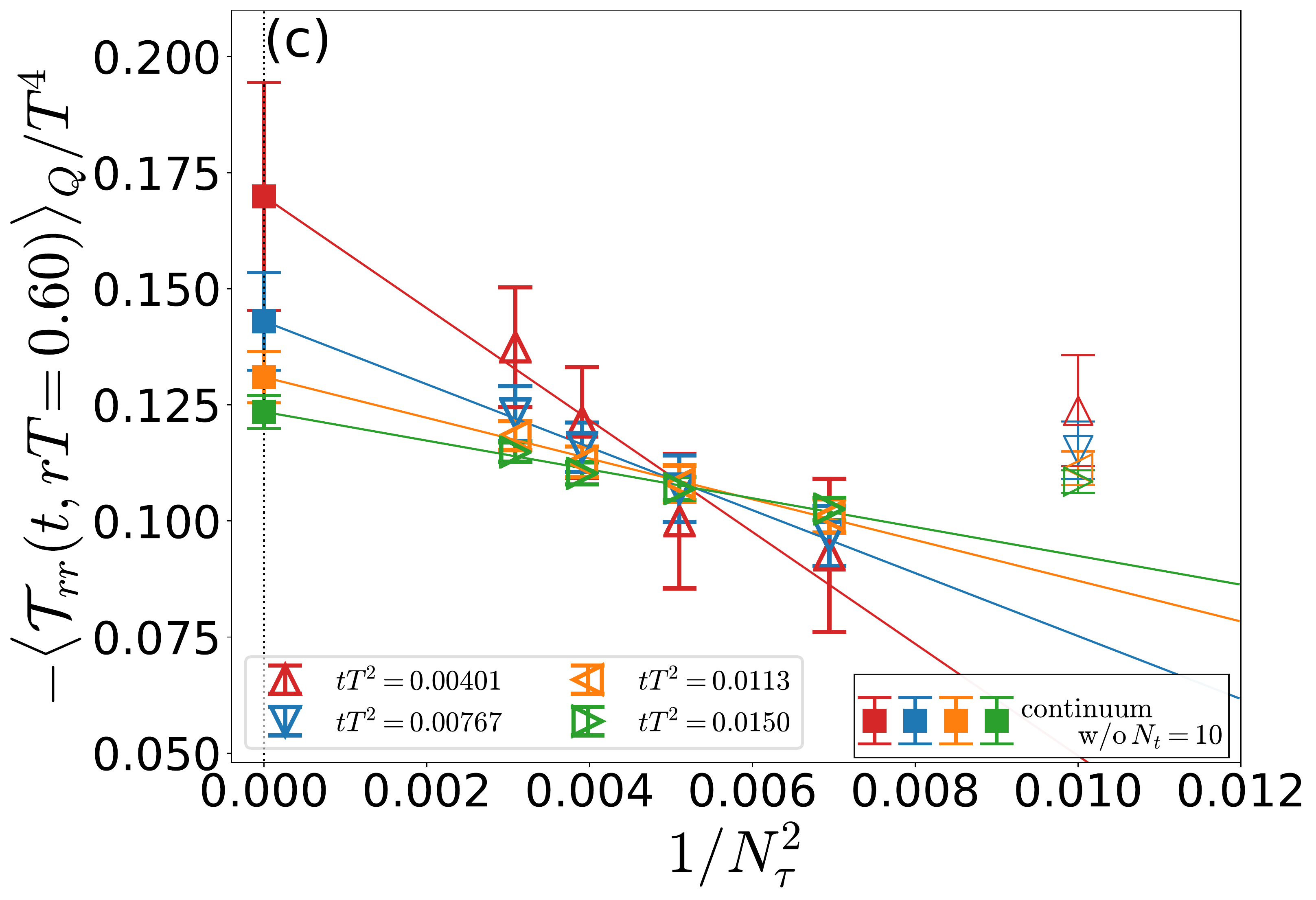}
 \caption{
 Color open symbols represent 
 $-\langle \mathcal{T}_{rr}(t, rT) \rangle_Q/T^4$
 for various $t$
 as functions of $1/N_\tau^2=a^2T^2$ at $T/T_c=1.44$.
 The continuum extrapolation for each $t$ is shown by the solid lines,
 with the extrapolated results represented by the filled symbols.
 Panels (a), (b) and (c) show the results at
 $rT=0.40,0.48,0.60$, respectively.
 }
 \label{fig:cont_lim}
\end{figure}

\begin{figure*}
 \centering
 \includegraphics[width=0.32\textwidth, clip]{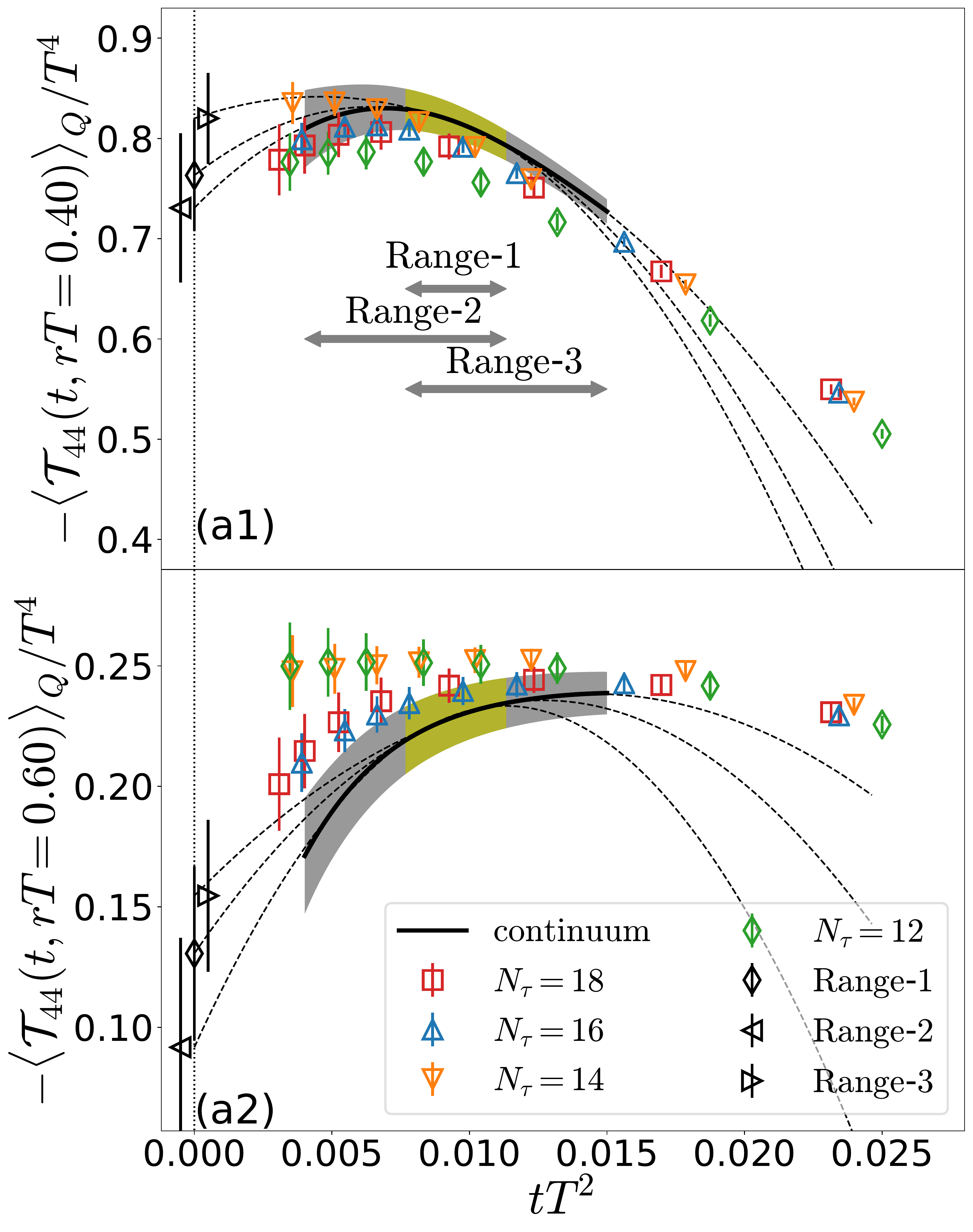}
 \includegraphics[width=0.32\textwidth, clip]{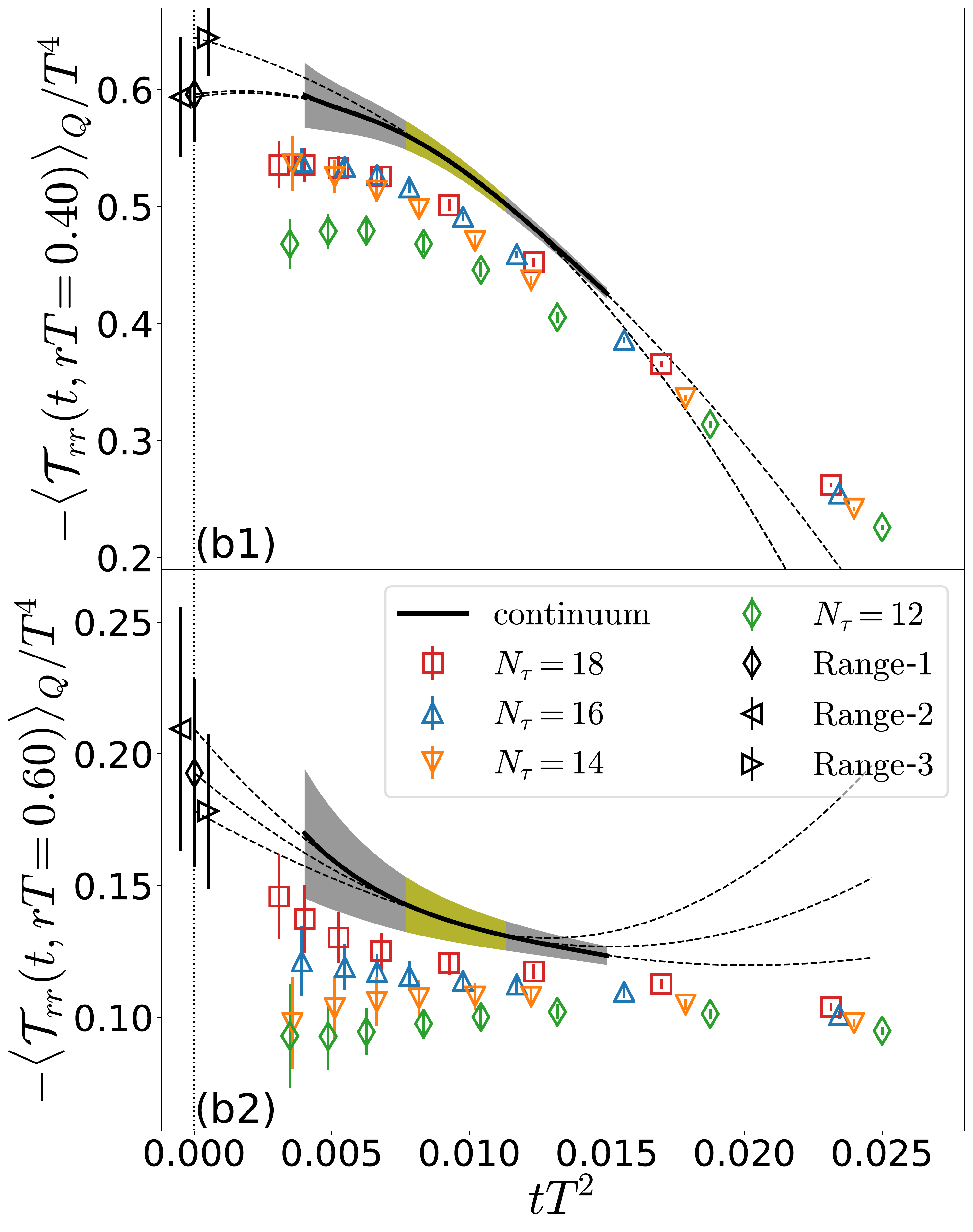}
 \includegraphics[width=0.32\textwidth, clip]{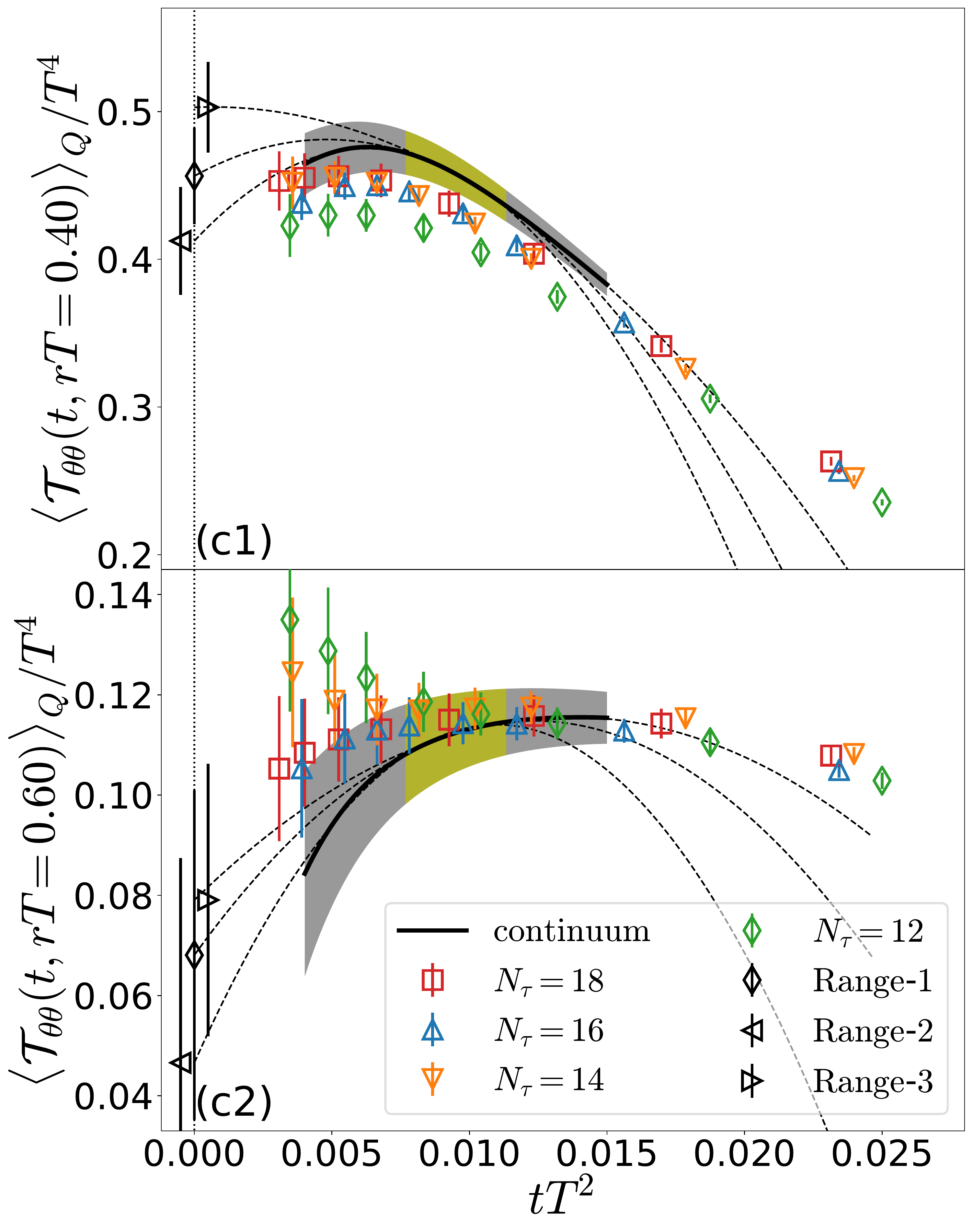}
 \caption{
 Each component of EMT at $rT=0.40$ (top) and $rT=0.60$ (bottom)
 as functions of $tT^2$.
 Color open symbols denote 
 $\langle \mathcal{T}_{\gamma\gamma'}(t, rT) \rangle_Q/T^4$
 for each $a$ as functions of $tT^2$. 
 The black solid line with the 
 gray error band is the continuum-extrapolated result.
 The dotted lines show the fitted results of the continuum result
 with Range-1, 2, and 3.
 The black symbols 
 are the results of the $t\to0$ extrapolation for these fitting ranges.
 }
 \label{fig:t_to_0}
\end{figure*}

In Fig.~\ref{fig:cont_lim}, we show 
$-\langle \mathcal{T}_{rr}(t, rT) \rangle_Q/T^4$
at $rT=0.40,0.48,0.60$ as a function of $1/N_\tau^2=(aT)^2$ 
for four values of $tT^2$.
To obtain the EMT values at a given $r$ and $t$, 
we first perform the linear interpolation of the lattice data
along the  $r$ direction and then interpolate 
along the $t$ direction by the cubic
spline method for each $a$.

In Fig.~\ref{fig:cont_lim},  
fitting results of the data at $N_\tau=12$--$18$
according to Eq.~(\ref{eq:double_lim}) at fixed $t$ 
are  shown by the solid lines,
while the results of the continuum limit are shown by 
the filled squares on the vertical dotted line at $1/N_\tau^2=0$.
We then take $t\rightarrow0$ extrapolation
by fitting the continuum-extrapolated results 
for different $t$ with Eq.~(\ref{eq:double_lim}) at $a=0$.
This fit has to be carried out within the range of $t$ satisfying 
Eq.~(\ref{eq:flow_require}).
We employ $tT^2=0.00401$ corresponding to
$t/a^2=1.3$ of the finest lattice data 
as the lower bound of the fitting window:
This choice satisfies Eq.~(\ref{eq:flow_require}) for all the lattices.
The upper bound of the fitting window is taken to be $tT^2=0.015$, since 
the thermodynamic quantities show a linear behavior
below this value ~\cite{Kitazawa:2016dsl}.

We consider the following three ranges  within
$0.00401 \leq tT^2 \leq 0.015$ to
estimate the systematic uncertainty  
from the fitting ranges~\cite{Kitazawa:2016dsl}:
\begin{align}
 &\mathrm{Range\mathchar`-1}:0.00767 \leq tT^2 \leq 0.0113, \notag
 \\
 &\mathrm{Range\mathchar`-2}:0.00401 \leq tT^2 \leq 0.0113, \notag
 \\
 &\mathrm{Range\mathchar`-3}:0.00767 \leq tT^2 \leq 0.0150. \notag
\end{align}
Range-1 is the most conservative window,
while Range-2 (Range-3) is the extension of 
Range-1 towards the smaller (larger) values of $t$.
We employ the result of Range-1 as a central value
and use the Range-2 and Range-3 for an estimate of the systematic error.
In the following, all the results after the double extrapolation
contain both the statistical and systematic errors.

In Fig.~\ref{fig:t_to_0}, the open symbols with statistical errors
represent
$\langle \mathcal{T}_{\gamma\gamma'}(t, rT) \rangle_Q/T^4$ 
at $rT=0.40$ (upper) and $0.60$ (lower) for each $a$.
The results of the continuum limit 
are denoted by the black solid lines with the gray
statistical error band
for $0.00401 \leq tT^2 \leq 0.015$.
Range-1 is highlighted by the yellow band.
The figure also shows the fitted results for Ranges-1, 2, and 3 
by the dotted lines.
The final results of the $t\to0$ limit for each range are shown by
the open black symbols around $tT^2=0$:
They agree with each other within the statistical errors,
which suggests that the systematic uncertainty from the choice of
the fitting range is not significant.

\section{Results of EMT distributions}
\label{sec:result}

Before entering into the detailed discussions on the 
spatial distribution of EMT, we first show  
the result of the stress distribution at $T/T_c = 2.60$
on a two-dimensional plane including the static source
in Fig.~\ref{fig:stress_around_quark}.
The same result is later shown in Fig.~\ref{fig:each_temp_r4}
in a different form.
In Fig.~\ref{fig:stress_around_quark},
the red and blue arrows represent the principal directions of the
stress tensor along the radial and transverse directions, respectively.
The length of each arrow represents square root of the eigenvalue
corresponding to each principal axis. This figure is to be compared with
Fig.~1 in Ref.~\cite{Yanagihara:2018qqg}.

\subsection{Channel dependence}

\begin{figure}
 \centering
 \includegraphics[width=0.36\textwidth, clip]{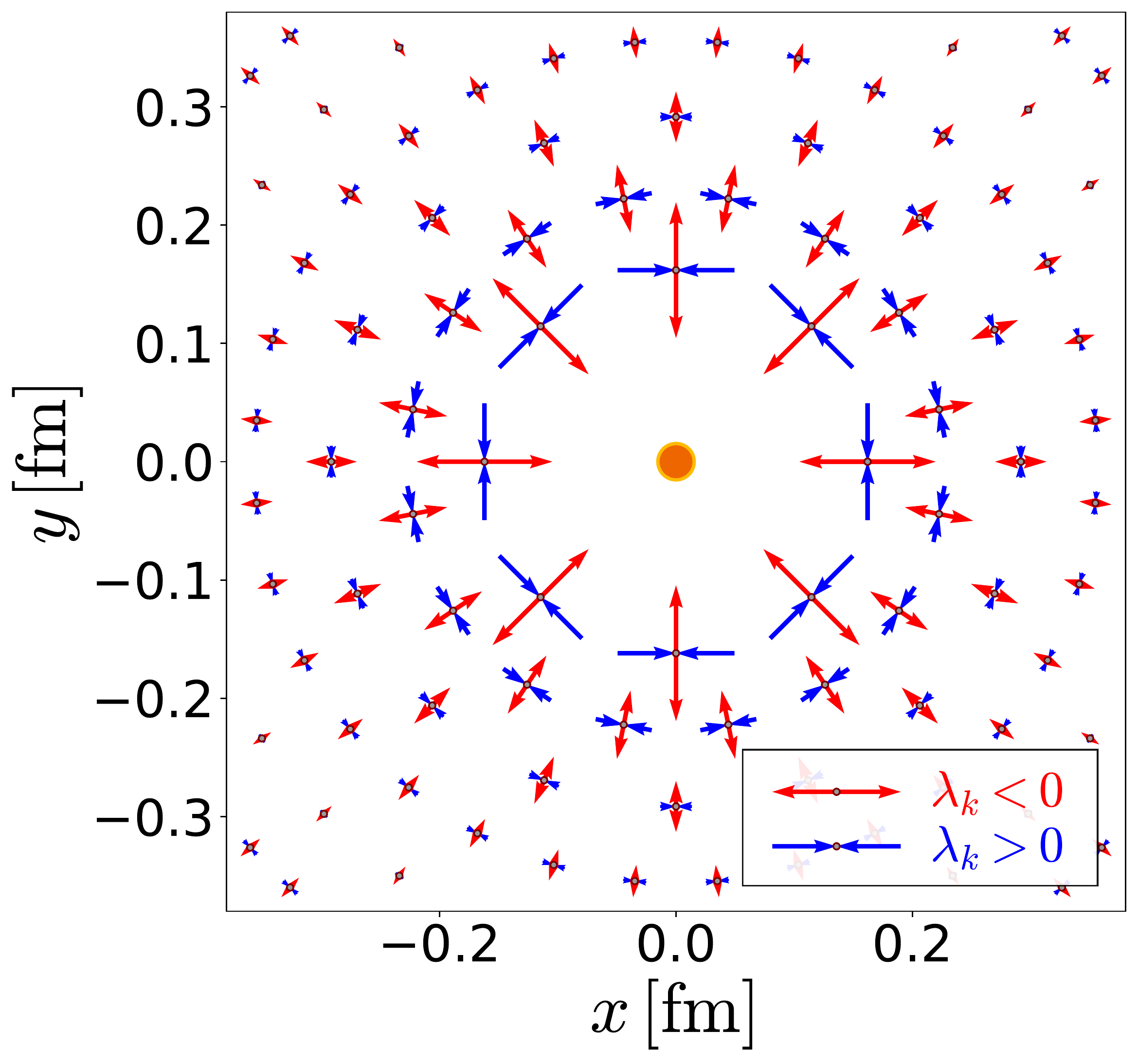}
 \caption{
 Stress distribution around a static quark
 at the origin at $T/T_c=2.60$.
 The red and blue arrows are the principal directions 
 along the radial and transverse directions, respectively.
 The length of each arrow represents square root of the eigenvalue.
 }
 \label{fig:stress_around_quark}
\end{figure}

\begin{figure*}
 \centering
 \includegraphics[width=0.32\textwidth, clip]{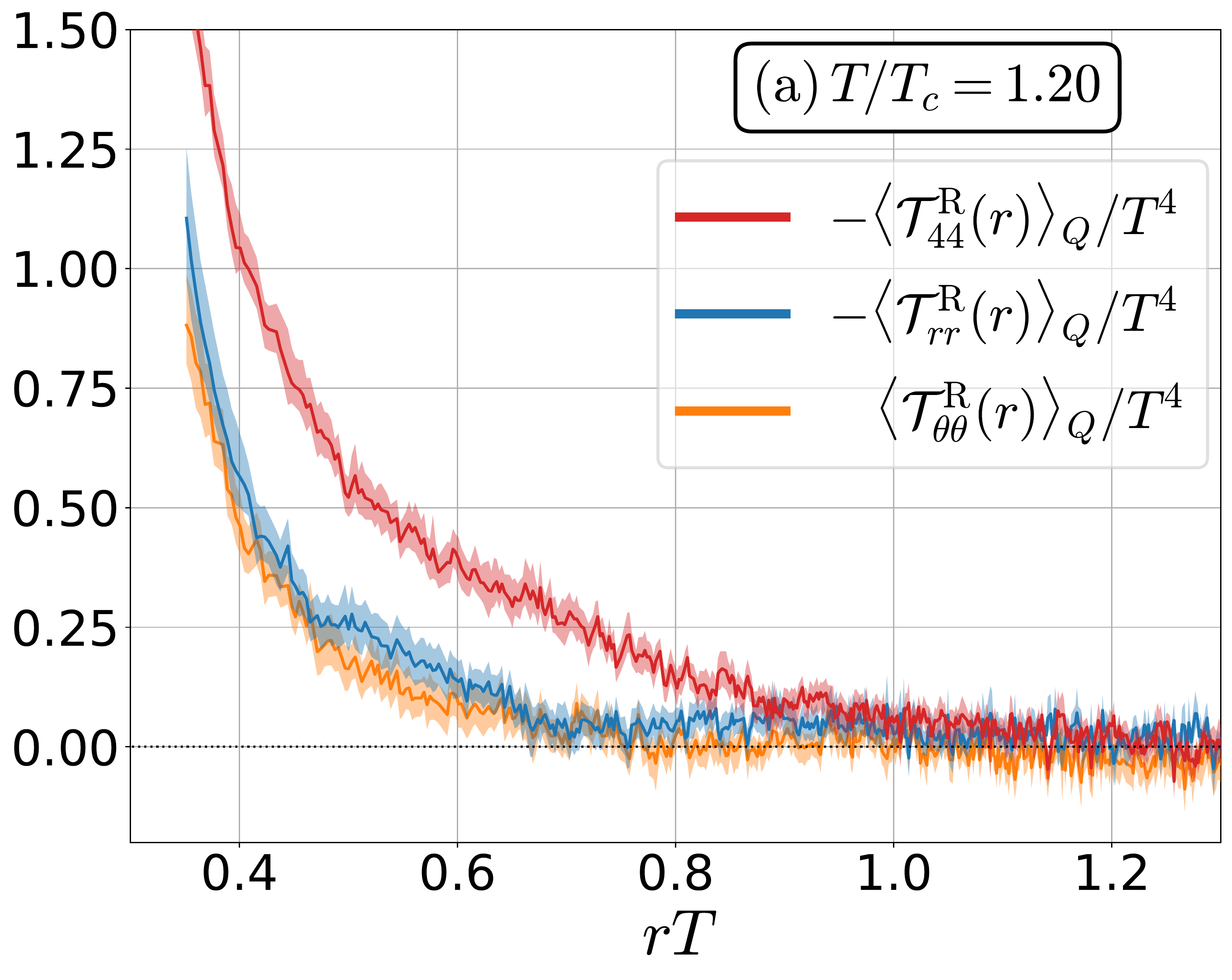}
 \includegraphics[width=0.32\textwidth, clip]{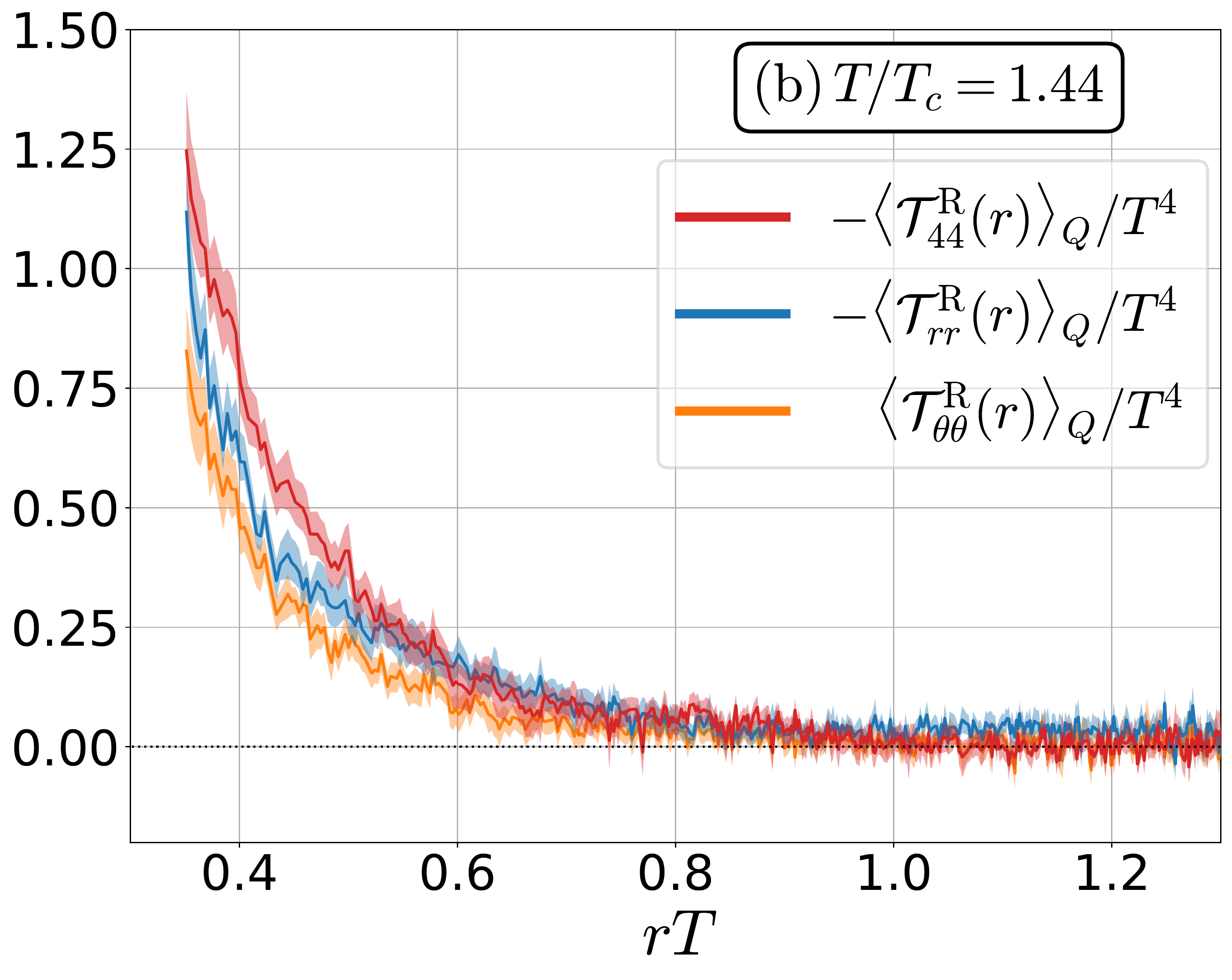}
 \\
 \includegraphics[width=0.32\textwidth, clip]{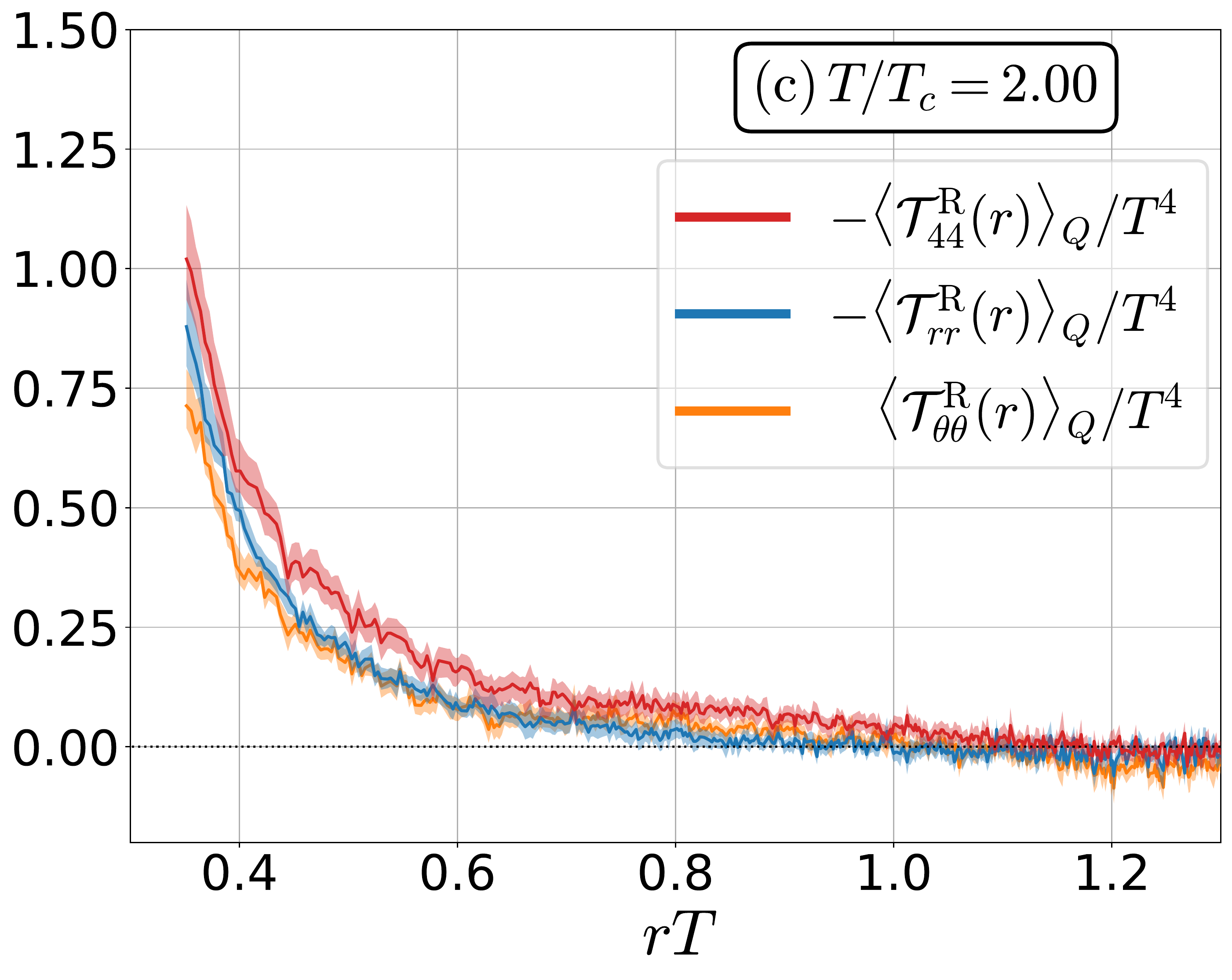}
 \includegraphics[width=0.32\textwidth, clip]{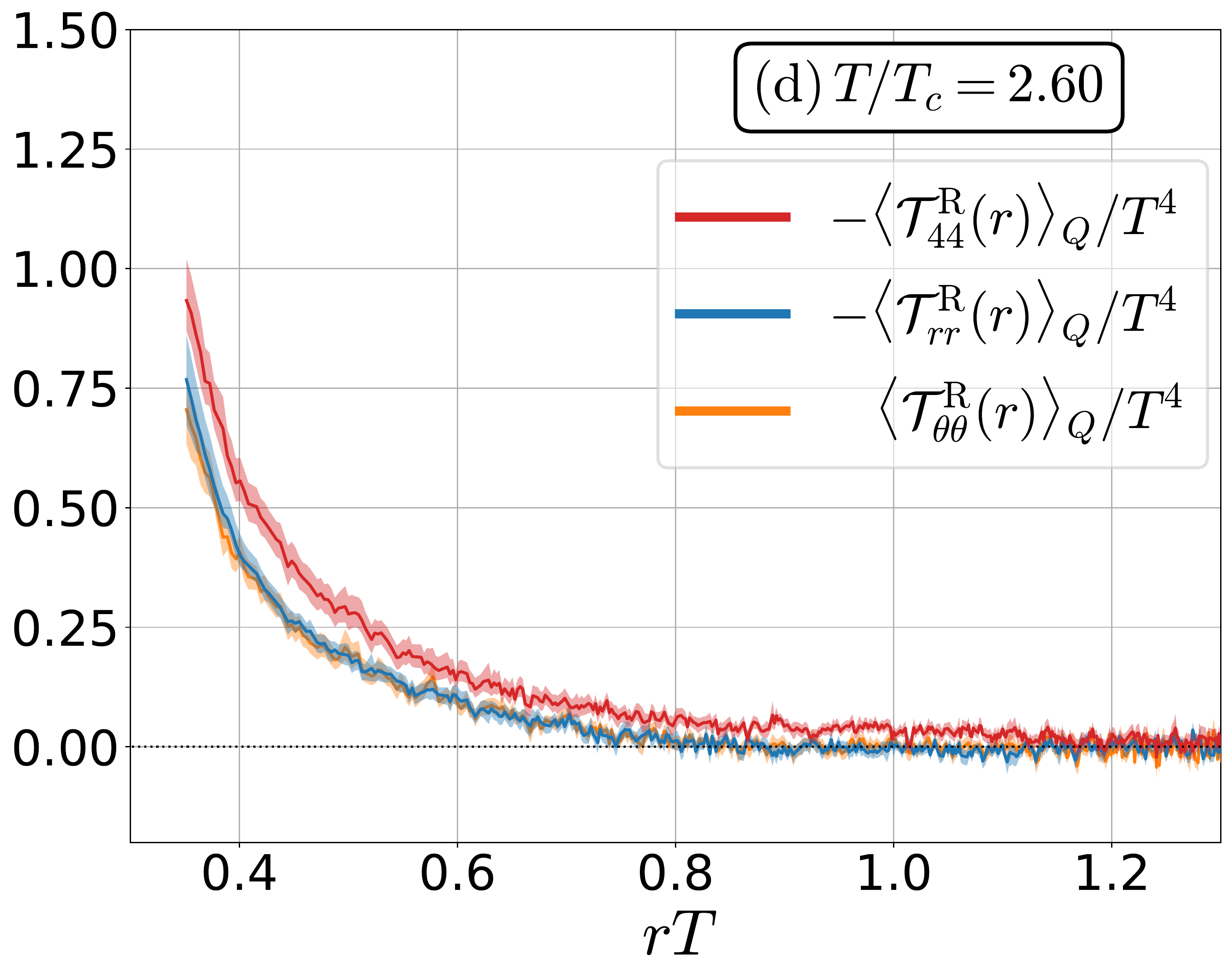}
 \caption{
 EMT distribution 
 $( -\langle{\cal T}_{44}^\mathrm{R} (r)\rangle_{Q},
 -\langle {\cal T}_{rr}^\mathrm{R} (r)\rangle_{Q},
 \langle {\cal T}_{\theta \theta}^\mathrm{R} (r)\rangle_{Q} )$
 as functions of $rT$
 after the double extrapolation:
 (a) $T=1.20T_c$,
 (b) $T=1.44T_c$,
 (c) $T=2.00T_c$, and
 (d) $T=2.60T_c$.
 }
 \label{fig:each}
\end{figure*}

\begin{figure}[t]
 \centering
 \includegraphics[width=0.32\textwidth, clip]{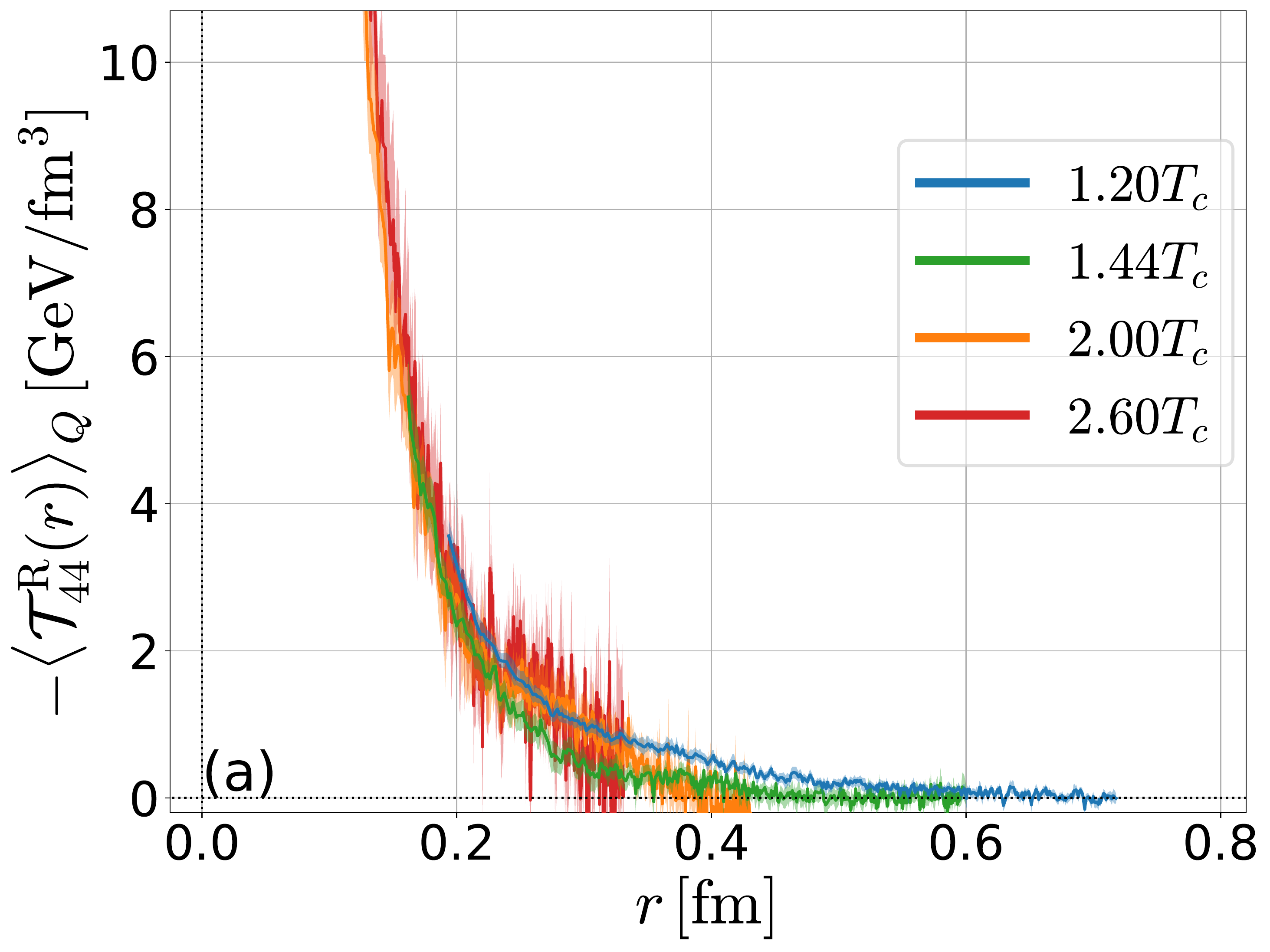}
 \includegraphics[width=0.32\textwidth, clip]{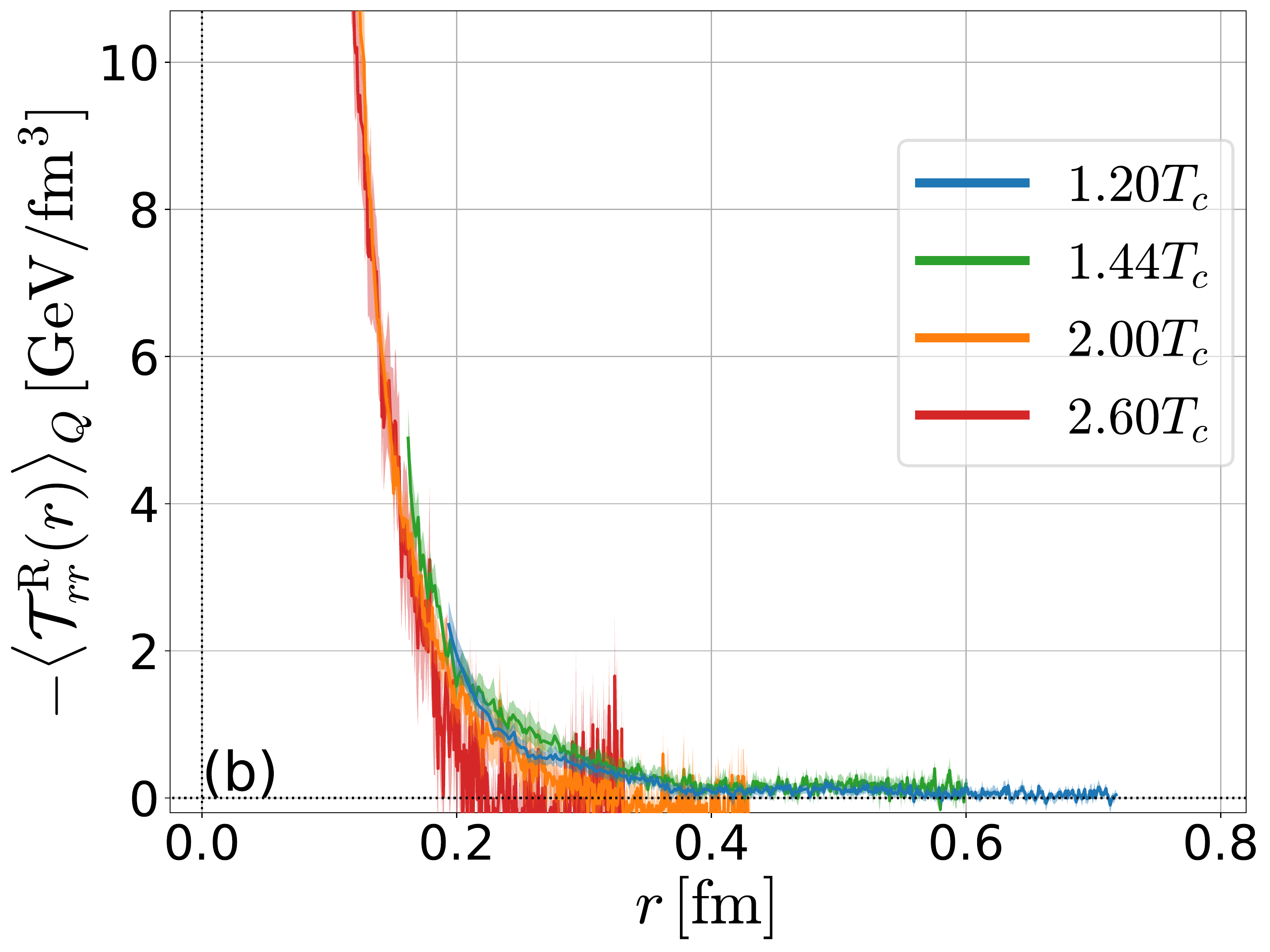}
 \\
 \includegraphics[width=0.32\textwidth, clip]{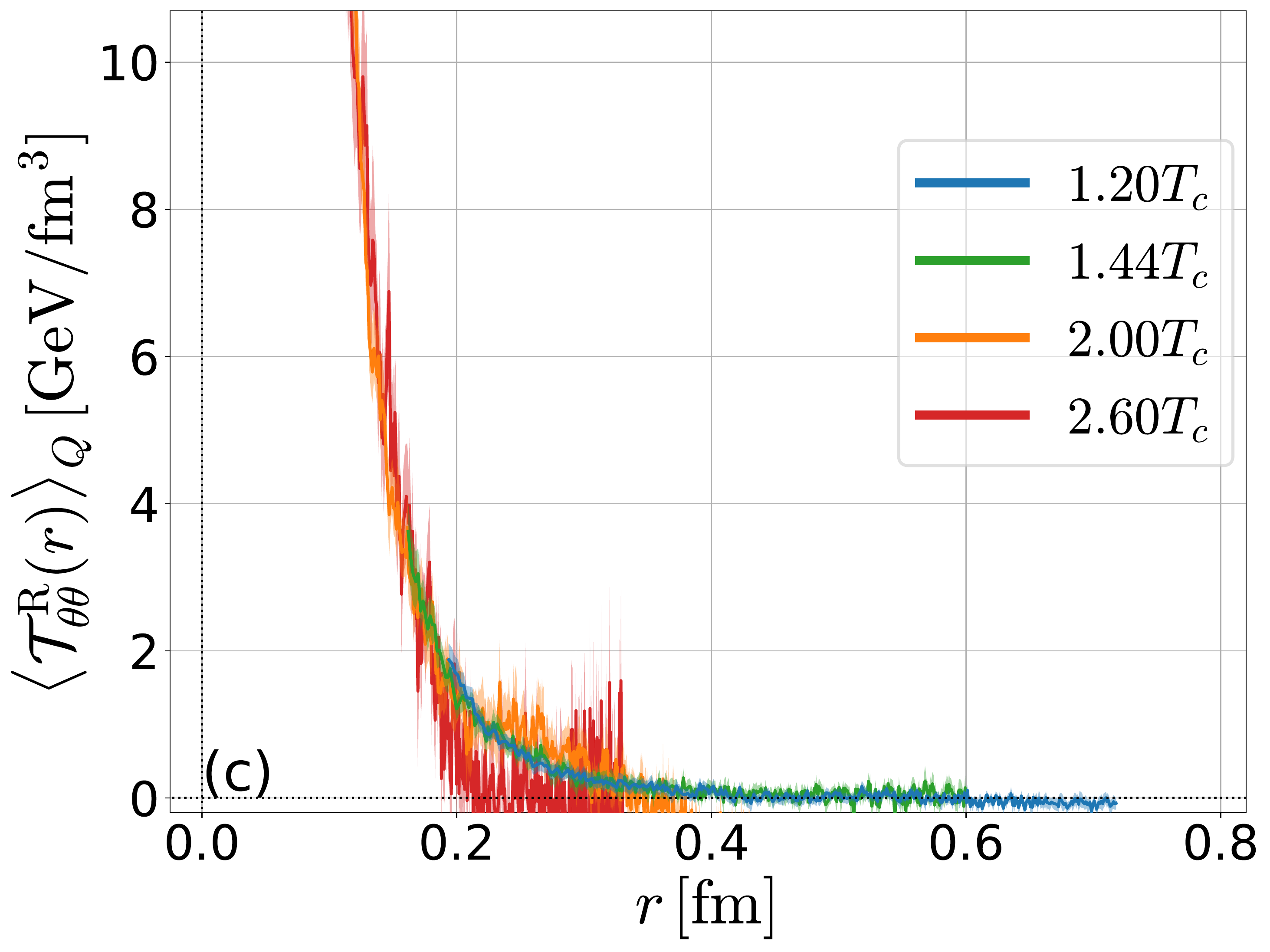}
 \includegraphics[width=0.32\textwidth, clip]{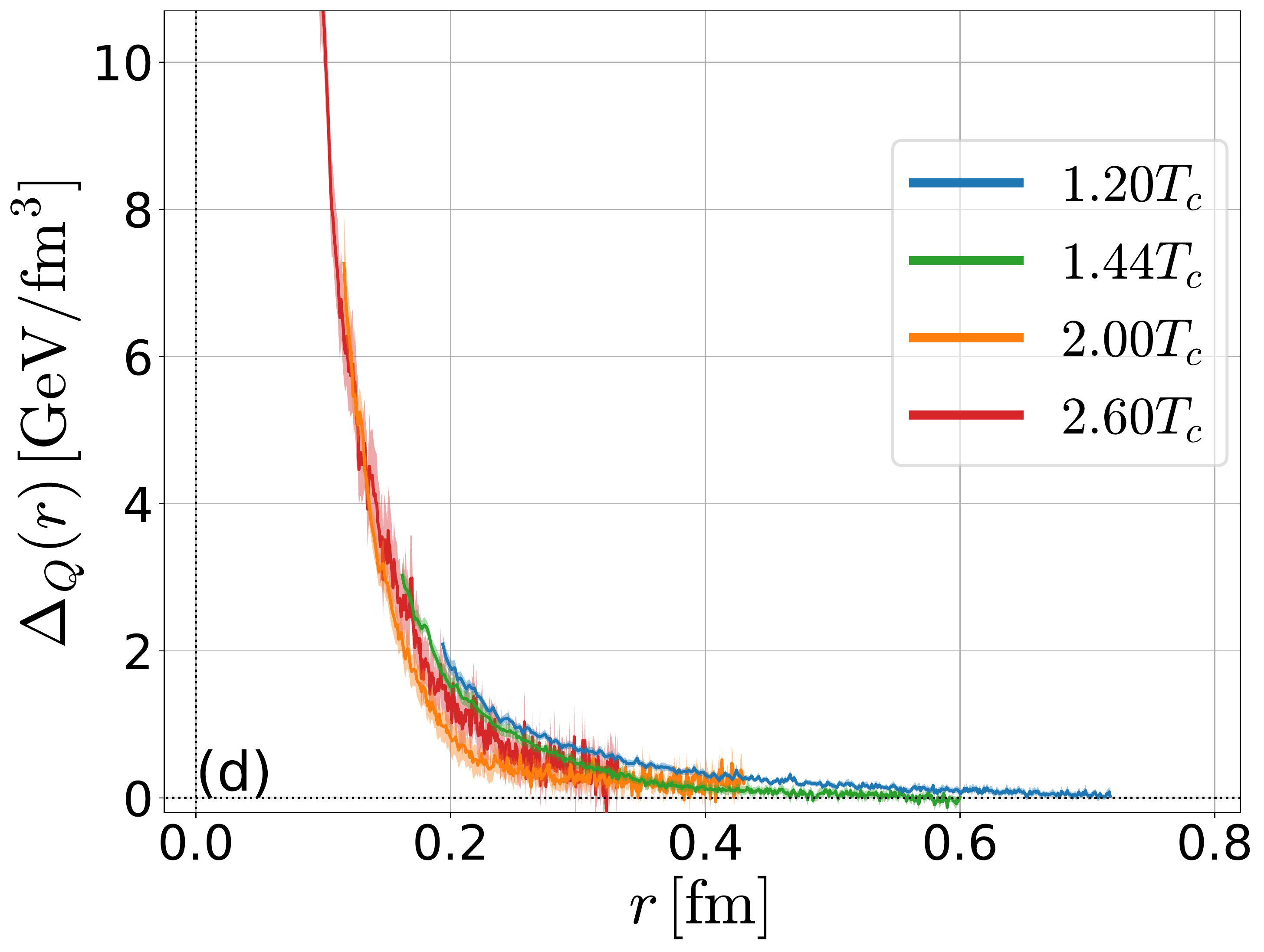}
 \caption{
 EMT distribution 
 $( -\langle{\cal T}_{44}^\mathrm{R} (r)\rangle_{Q},
 -\langle {\cal T}_{rr}^\mathrm{R} (r)\rangle_{Q},
 \langle {\cal T}_{\theta \theta}^\mathrm{R} (r)\rangle_{Q} )$
 and $\Delta_{Q}(r)$
 as functions of $r$~[fm]
 at each temperature:
 (a) $-\langle{\cal T}_{44}^\mathrm{R} (r)\rangle_{Q}$,
 (b) $-\langle{\cal T}_{rr}^\mathrm{R} (r)\rangle_{Q}$,
 (c) $\langle{\cal T}_{\theta \theta}^\mathrm{R} (r)\rangle_{Q}$, and
 (d) $\Delta_{Q}(r)$.
 }
 \label{fig:each_temp}
\end{figure}

\begin{figure}[t]
 \centering
 \includegraphics[width=0.34\textwidth, clip]{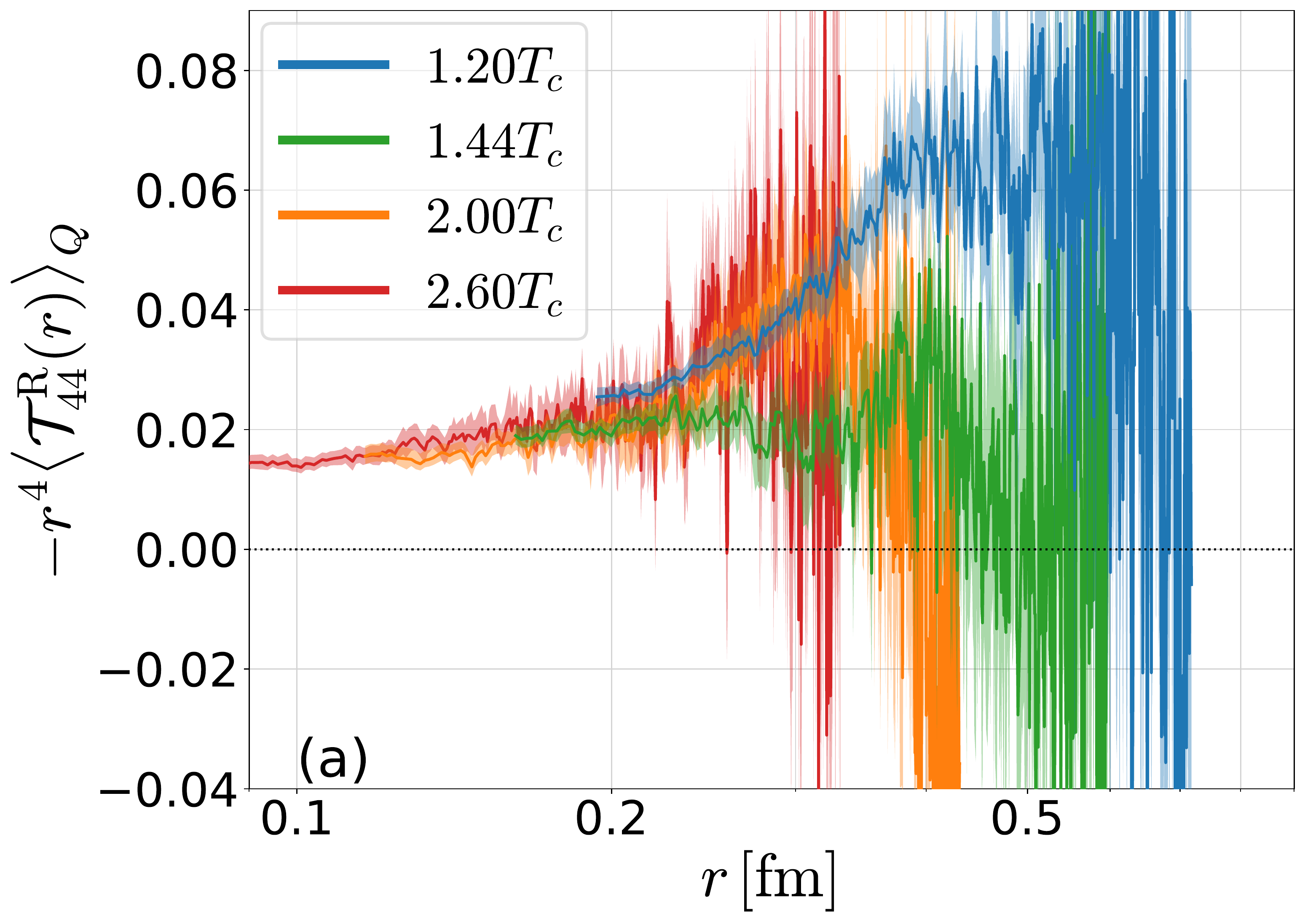}  
 \includegraphics[width=0.34\textwidth, clip]{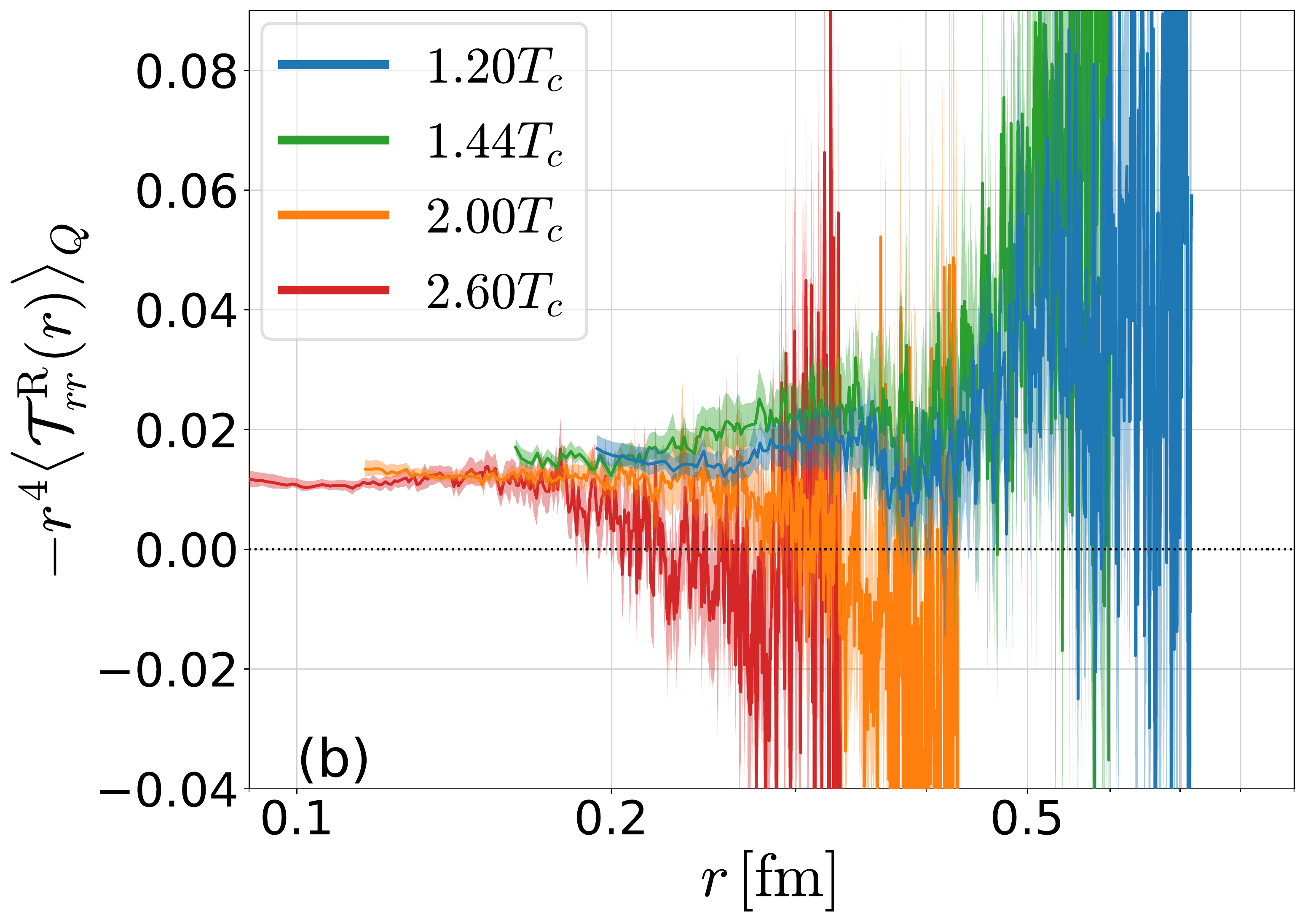}
 \\
 \includegraphics[width=0.34\textwidth, clip]{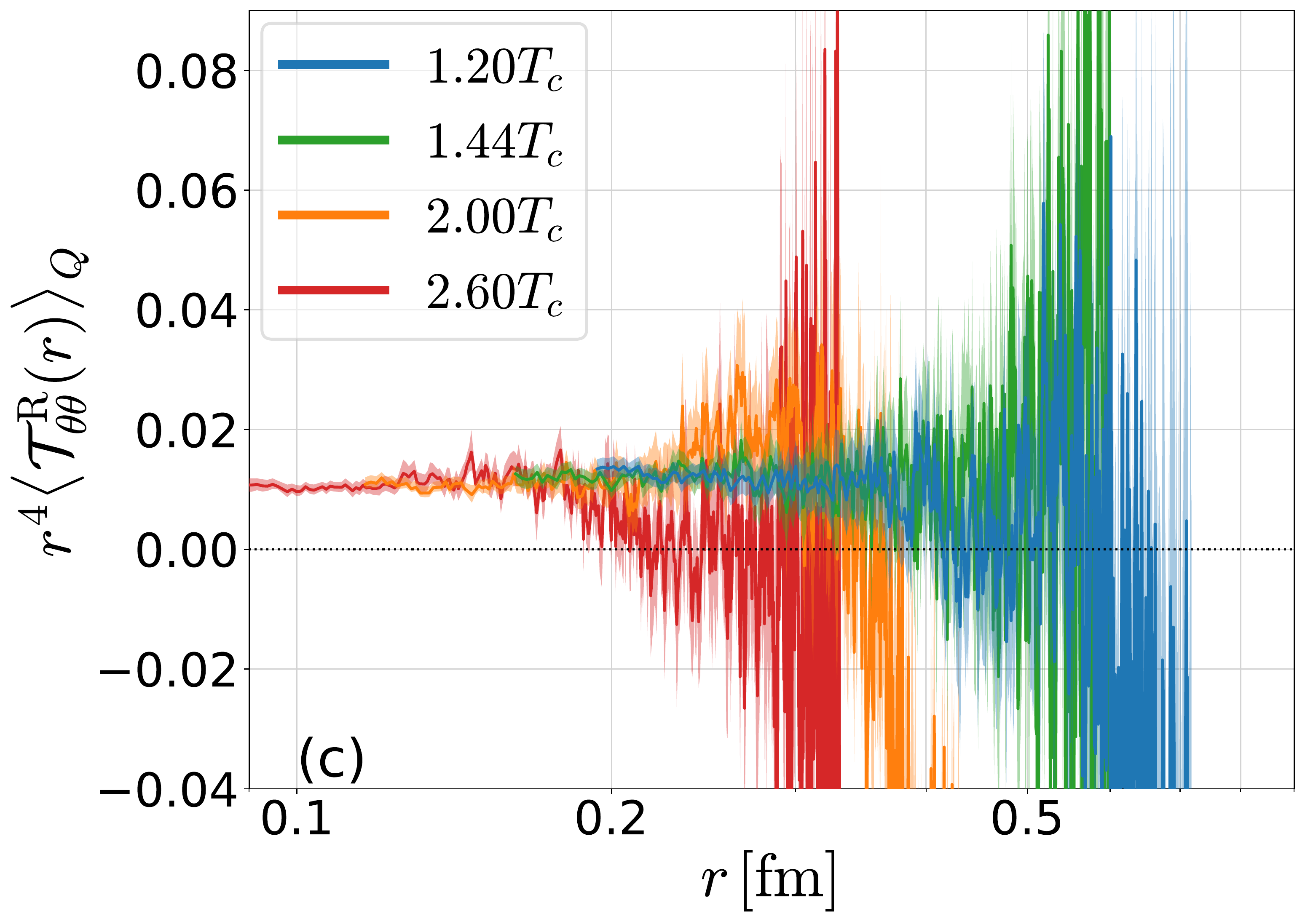}
 \includegraphics[width=0.34\textwidth, clip]{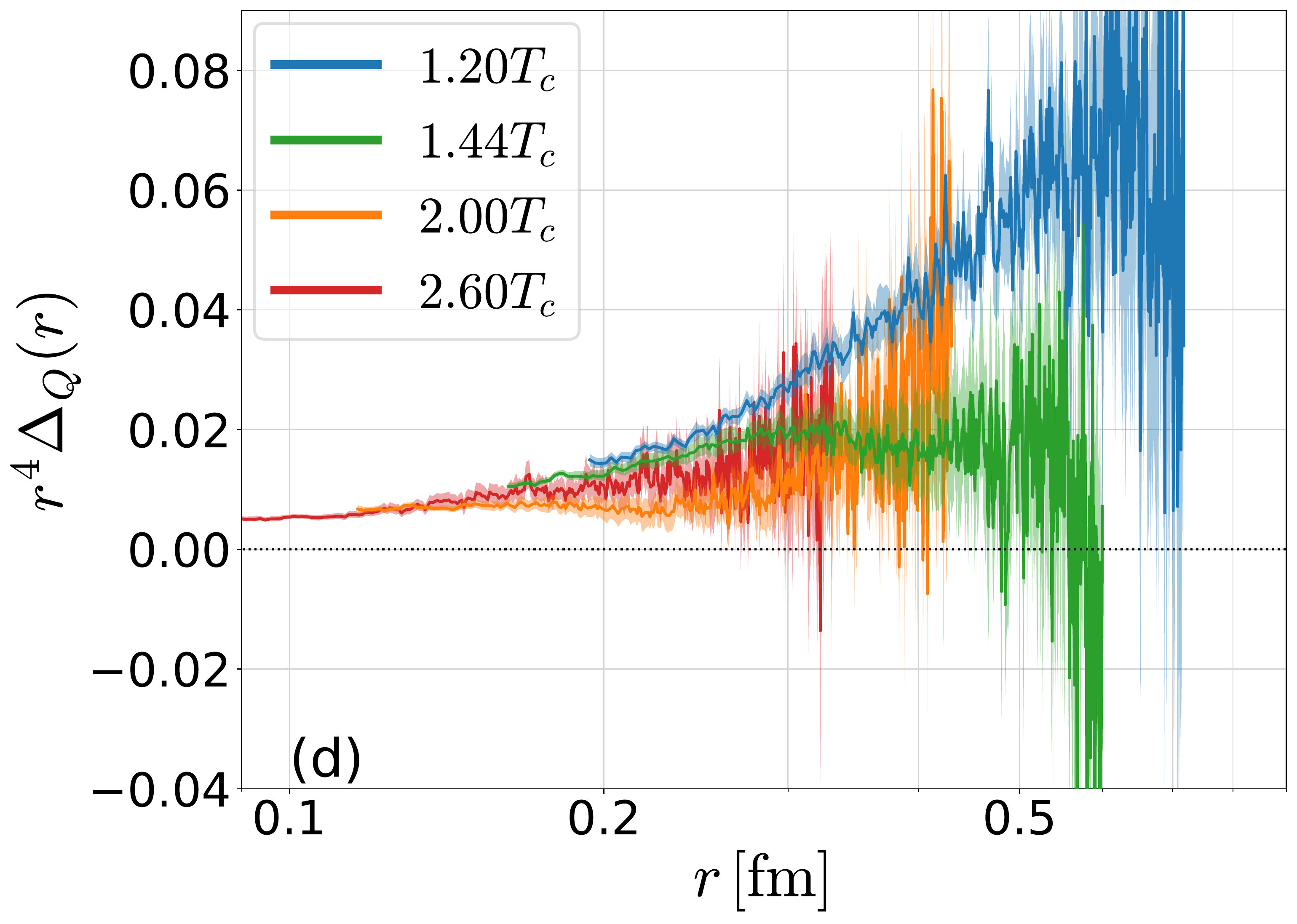}
 \caption{
 EMT distribution 
 $r^4( -\langle{\cal T}_{44}^\mathrm{R} (r)\rangle_{Q},
 -\langle {\cal T}_{rr}^\mathrm{R} (r)\rangle_{Q},
 \langle {\cal T}_{\theta \theta}^\mathrm{R} (r)\rangle_{Q} )$
 and $r^4\Delta_{Q}(r)$
 as functions of $r$~[fm]
 at each temperature:
 (a)$-r^4\langle{\cal T}_{44}^\mathrm{R} (r)\rangle_{Q}$,
 (b)$-r^4\langle{\cal T}_{rr}^\mathrm{R} (r)\rangle_{Q}$,
 (c)$r^4\langle{\cal T}_{\theta \theta}^\mathrm{R} (r)\rangle_{Q}$, and
 (d)$r^4\Delta_Q (r)$.
 }
 \label{fig:each_temp_r4}
\end{figure}

In Fig.~\ref{fig:each}, we show the dimensionless EMT, 
$-\langle{\cal T}_{44}^\mathrm{R} (r)\rangle_{Q}/T^4$,
$-\langle {\cal T}_{rr}^\mathrm{R} (r)\rangle_{Q}/T^4$, and 
$\langle {\cal T}_{\theta \theta}^\mathrm{R} (r)\rangle_{Q} /T^4$,
as functions of the dimensionless length $rT$. 
The error bands include both the statistical and systematic errors,
where the latter is estimated from the three fitting ranges for
the $t\to0$ extrapolation.
Since the thermal expectation value
$\langle \mathcal{T}_{\mu\nu}(t,x)\rangle$ is subtracted 
as in Eq.~(\ref{eq:cor_pol_emt0}), we have 
$\langle \mathcal{T}^\mathrm{R}_{\mu\nu}(r)\rangle_Q\to0$ 
in the $r\to\infty$ limit.

We find that  $-\langle{\cal T}_{44}^\mathrm{R} (r)\rangle_{Q}$, 
$-\langle {\cal T}_{rr}^\mathrm{R} (r)\rangle_{Q}$, and 
$\langle {\cal T}_{\theta \theta}^\mathrm{R} (r)\rangle_{Q}$ 
are all positive for $rT\lesssim1$ and
decrease rapidly with increasing $r$.
These signs are the same as those of 
the Maxwell stress tensor in Eq.~(\ref{eq:Maxwell}).
Individual signs physically mean that a volume element 
has a positive localized energy density and receives
a pulling (pushing) force along the longitudinal (transverse)
direction;
see Figs.~\ref{fig:stress_pic} and \ref{fig:stress_around_quark}.

Figure~\ref{fig:each} indicates that 
the absolute values of the spatial components 
$|\langle {\cal T}_{rr}^\mathrm{R} (r)\rangle_{Q}|$ and 
$|\langle {\cal T}_{\theta \theta}^\mathrm{R} (r)\rangle_{Q}|$
are degenerated within the error for 
all temperatures.
On the other hand, 
$|\langle{\cal T}_{44}^\mathrm{R}  (r)\rangle_{Q}|$ is larger than 
the spatial components
especially at lower temperature.
This is in contrast to the degenerate magnitude of all components 
in  the Maxwell stress Eq.~(\ref{eq:Maxwell}) and is also 
different from the leading-order thermal perturbation theory 
(Appendix~\ref{sec:pert_calc}).

\subsection{Temperature dependence}

Shown in Fig.~\ref{fig:each_temp} is
the temperature dependence of 
the spatial distribution of EMT  
with respect to the physical distance $r$~[fm];
(a) $-\langle{\cal T}_{44}^\mathrm{R} (r)\rangle_{Q}$,
(b) $-\langle {\cal T}_{rr}^\mathrm{R} (r)\rangle_{Q}$,
and 
(c) $\langle {\cal T}_{\theta \theta}^\mathrm{R} (r)\rangle_{Q} $.
Also, shown in Fig.~\ref{fig:each_temp}(d) is the distribution of 
the trace of EMT given by
\begin{align}
 \Delta_Q(r) &\equiv -\langle \mathcal{T}^\mathrm{R}_{\mu\mu}(r) \rangle_Q 
 \notag
 \\
 &= -\langle \mathcal{T}^\mathrm{R}_{44}(r)
 + \mathcal{T}^\mathrm{R}_{rr}(r) 
 + 2\mathcal{T}^\mathrm{R}_{\theta \theta}(r)\rangle_Q.
\end{align}
Figure~\ref{fig:each_temp} tells us that
the EMT distributions  have small $T$ dependence at  
short distances, $r \lesssim 0.2$ fm.
On the other hand, for large distances, sizable $T$ dependence can be seen 
despite the growth of the errors at high $T$.

To make these features more explicit, we plot the same results
with a dimensionless normalization
$r^4 \langle{\cal T}_{\gamma \gamma'}^\mathrm{R} (r)\rangle_{Q}$ 
as a function of $r$
in Fig.~\ref{fig:each_temp_r4}.
The figure shows
that the $T$ dependence is suppressed for
$rT \lesssim 0.3$ and all results approach a single line, while
the result
tends to be more suppressed for
$rT \gtrsim 0.3$ compared with this universal behavior
as temperature is raised.
This result is reasonable as the $T$ dependence of 
$r^4 \langle{\cal T}_{\gamma \gamma'}^\mathrm{R} (r)\rangle_{Q}$ would
be suppressed for $r\lesssim(2\pi T)^{-1}$.

At distance $(2\pi T)^{-1} \ll r\ll\Lambda^{-1}$ with
$\Lambda$ being the lambda parameter,
the behavior of $\langle{\cal T}_{\gamma \gamma'}^\mathrm{R}(r)\rangle_{Q}$
should be described by the perturbation theory in  electrostatic QCD (EQCD).
In the leading order of EQCD in this regime, we have the following ratio
(Appendix~\ref{sec:pert_calc}),
\begin{align}
 \left|\frac{\Delta_Q(r) }{ \langle {\cal T}^\mathrm{R}_{44,rr,\theta\theta}(r) \rangle_Q}\right|
 = \frac{11}{2\pi} \alpha_s + \mathcal{O}(g^3),
 \label{eq:alpha}
\end{align}
which is independent of $r$ and $T$ and 
is given only by a function of $\alpha_s$.

\begin{figure}[t]
 \centering
 \includegraphics[width=0.36\textwidth, clip]{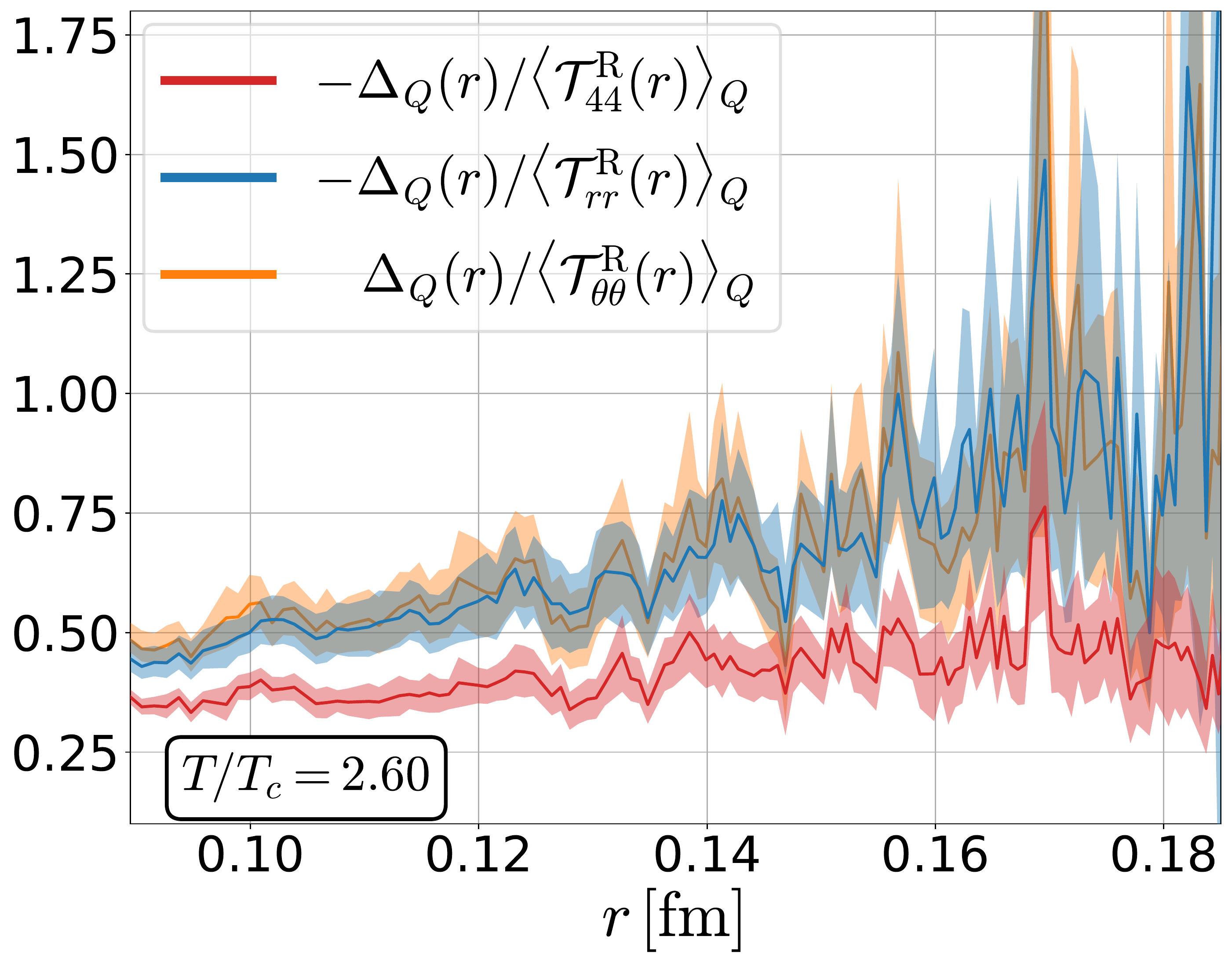}
 \caption{
 Ratios 
 $\Delta_Q(r) / \langle\mathcal{T}_{44,rr,\theta\theta}^\mathrm{R}(r) \rangle_Q$
 as functions of $r$ at $T/T_c=2.60$.
 }
 \label{fig:ratio_mu-4rt}
\end{figure}

Shown in Fig.~\ref{fig:ratio_mu-4rt} is the $r$ dependence of 
$\Delta_Q(r) / \langle\mathcal{T}_{44,rr,\theta\theta}^\mathrm{R}(r) \rangle_Q$
as a function of $r$ at $T/T_c=2.60$.
From this result and  Eq.~(\ref{eq:alpha}), we obtain, at $r=0.1$ fm, 
$\alpha_s = 0.221(17)$ 
from $-\Delta_Q(r) / \langle {\cal T}^\mathrm{R}_{44}(r) \rangle_Q$,
$\alpha_s = 0.286(24)$ 
from $-\Delta_Q(r) / \langle {\cal T}^\mathrm{R}_{rr}(r) \rangle_Q$,
and $\alpha_s = 0.319(35)$ 
from $\Delta_Q(r) / \langle {\cal T}^\mathrm{R}_{\theta\theta}(r)
\rangle_Q$.
Although these values are channel dependent, indicating the 
existence of non-negligible higher order contributions, it is notable
that they are consistent with that obtained from
the similar analysis of the Polyakov loop correlations
at $r=0.1$~fm~\cite{Kaczmarek:2004gv,Kaczmarek:2005ui}.   
Higher order $\alpha_s$ corrections and the 
thermal corrections for EMT around a static charge to be compared with our
lattice data  is under way \cite{Matthias}.

Let us now turn to the long-distance region 
in Fig.~\ref{fig:each_temp_r4}. 
Owing to the large errors in this region, it is not 
possible to extract the
thermal screening of the form $\exp (-2m_{\rm _D} r)$ 
with $m_{\rm _D}$ being the 
Debye screening mass. Nevertheless, Figs.~\ref{fig:each_temp_r4}(a-d)
indicate that the EMT distributions decrease
faster than $1/r^4$ at long distances, and the tendency is stronger 
at  high temperatures. 
To draw a definite conclusion, however, higher statistical data 
are necessary.

\section{Summary and Concluding remarks}
\label{sec:summary}

In the present paper, 
we have studied, for the first time, 
the EMT distribution around a static quark
at finite temperature above $T_c$ of the SU(3) YM theory on the lattice.
The YM gradient flow plays crucial roles to define the EMT
on the lattice and to explore its spatial structure. 

The main results of this paper can be summarized as follows.

As shown in Fig.~\ref{fig:each}, we found no significant 
difference between the absolute magnitude of EMT
along the radial direction and that of the transverse direction
for all temperatures above $T_c$. This seems to be
in accordance with the leading-order thermal perturbation 
theory in QCD, which predicts the same magnitude for 
all principal components of EMT. 
However,  we found a  substantial difference between 
the EMT distribution in the temporal direction and that of the 
spatial directions, especially near $T_c$.
This indicates that there is indeed a
genuine non-Abelian effect present at finite temperature,
so that precise comparison with the higher-order
thermal QCD calculation would be called for. 
  
As shown in  Figs.~\ref{fig:each_temp} and \ref{fig:each_temp_r4},
all the  EMT distributions  have small $T$ dependence at  
short distances, $r \lesssim 0.2$ fm.
Also the EMT distributions decrease
faster than $1/r^4$ at long distances, and the tendency is stronger 
at high temperatures. However, owing to the large statistical errors,
we could not extract the values of the thermal Debye screening.
By using the fact that the EMT distributions 
are $T$ independent at short distances,
we attempted to extract the strong coupling constant
from the  ratios between the different components of EMT.
The result, $\alpha_s(0.1\ {\rm fm}) \simeq  0.28-0.32$, is 
consistent with that  obtained from the similar analysis 
for the $Q\bar{Q}$ free energy at finite $T$.

We have  some important issues to be  studied further: 
Going beyond the leading-order 
thermal QCD calculation for the EMT ~\cite{Matthias} is necessary 
to understand  the lattice results presented in this paper. 
At the same time,  increasing  the statistics of lattice data
is  necessary to extract, e.g. the screening mass  from the 
long range part of the EMT distribution.

There are also several interesting future  problems.
First of all,  the extension to full QCD is an important  next step.
Since the $Z_3$ symmetry is explicitly broken by dynamical fermions,
the present method can be applied directly 
to a single static quark $Q$, a static diquark $QQ$ and $Q\bar{Q}$  
both at low and high temperatures.
In particular, the single quark system in QCD at zero temperature
corresponds to a heavy-light meson~\cite{Mueller:2019mkh}.
Secondly, the EMT distributions of the $QQQ$ system  
will provide new insight into the flux tube formation
in baryons~\cite{Takahashi:2002bw} 
as well as the ``gravitational'' baryon structure~\cite{Kumano:2017lhr,Polyakov:2018zvc,Burkert:2018bqq,Shanahan:2018nnv}
at zero and non-zero temperatures.

\section*{Acknowledgment}
The authors thank T.~Iritani for fruitful discussions 
in the early stage of this study.
They are also grateful to M.~Berwein for discussions regarding 
the perturbative analysis of the EMT distribution.
M.~K. thanks F.~Karsch for useful discussions.
The numerical simulation was carried out on OCTOPUS
at the Cybermedia Center, Osaka University and Reedbush-U
at Information Technology Center, The University of Tokyo.
This work was supported by JSPS Grant-in-Aid for Scientific Researches, 
17K05442, 18H03712, 18H05236, 18K03646, 19H05598, 20H01903.

\appendix

\section{Tree-level improvement of the lattice observables}
\label{sec:append_disc}

\begin{figure}[t]
 \centering
 \includegraphics[width=0.32\textwidth, clip]{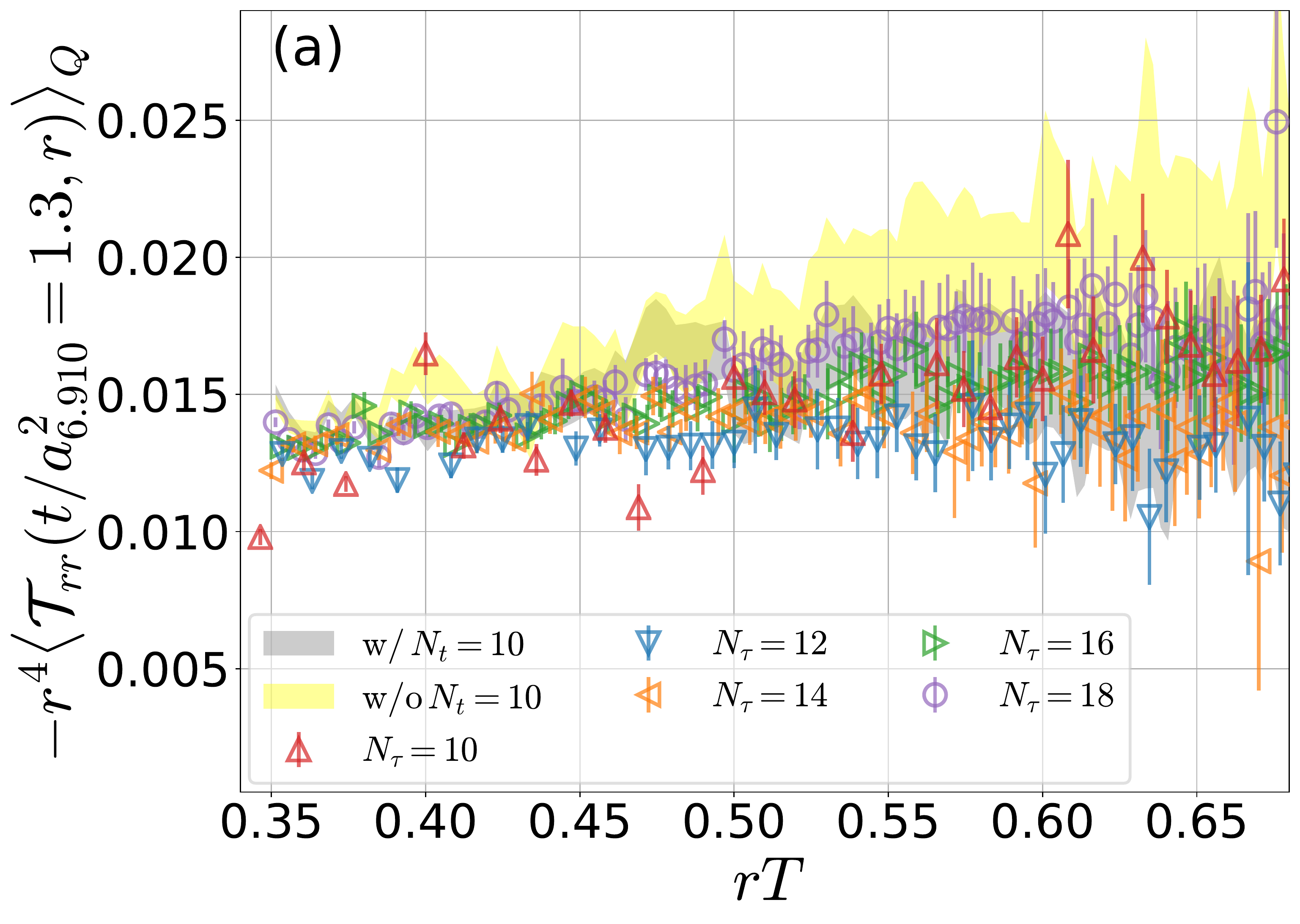}
 \includegraphics[width=0.32\textwidth, clip]{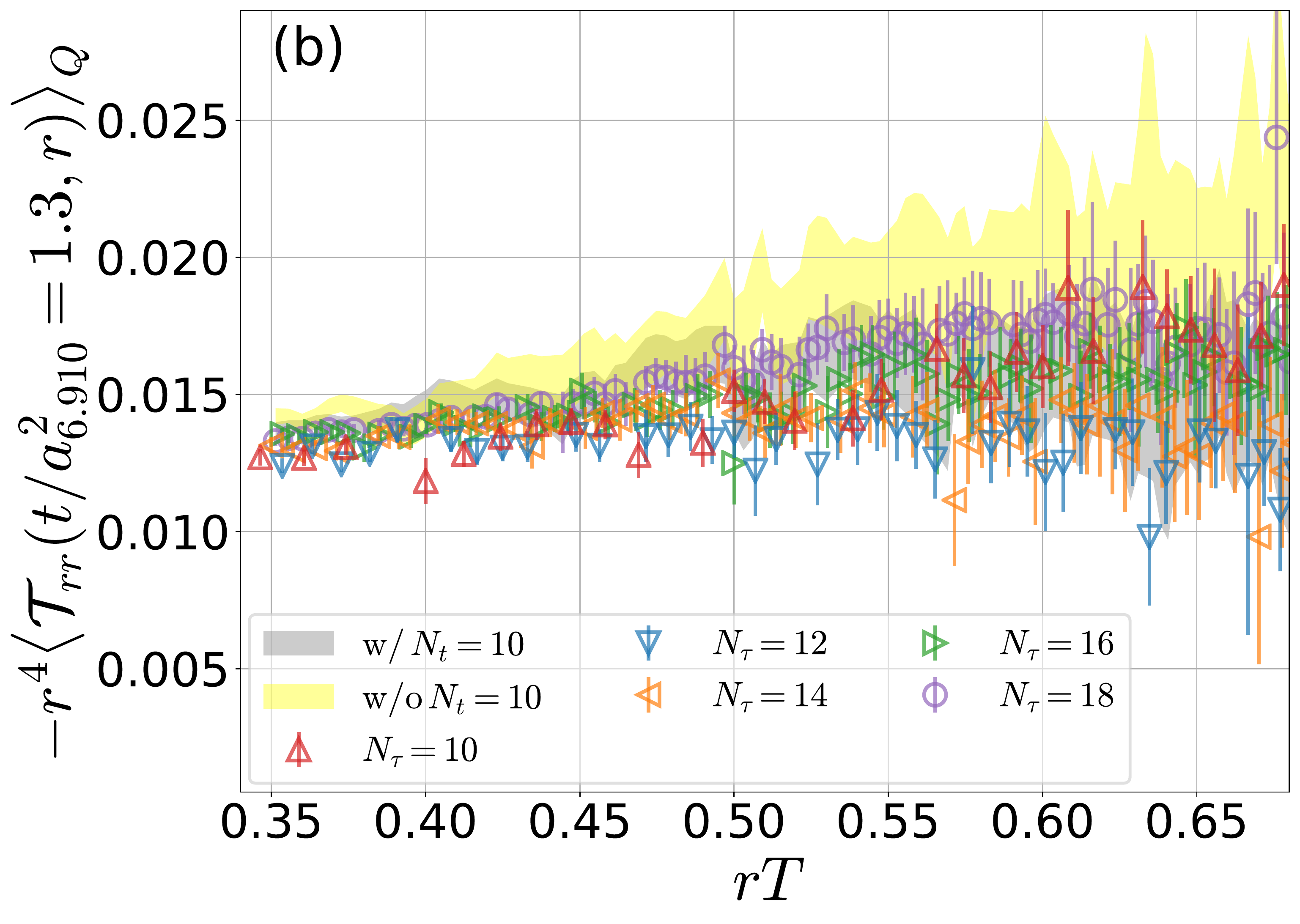}
 \includegraphics[width=0.32\textwidth, clip]{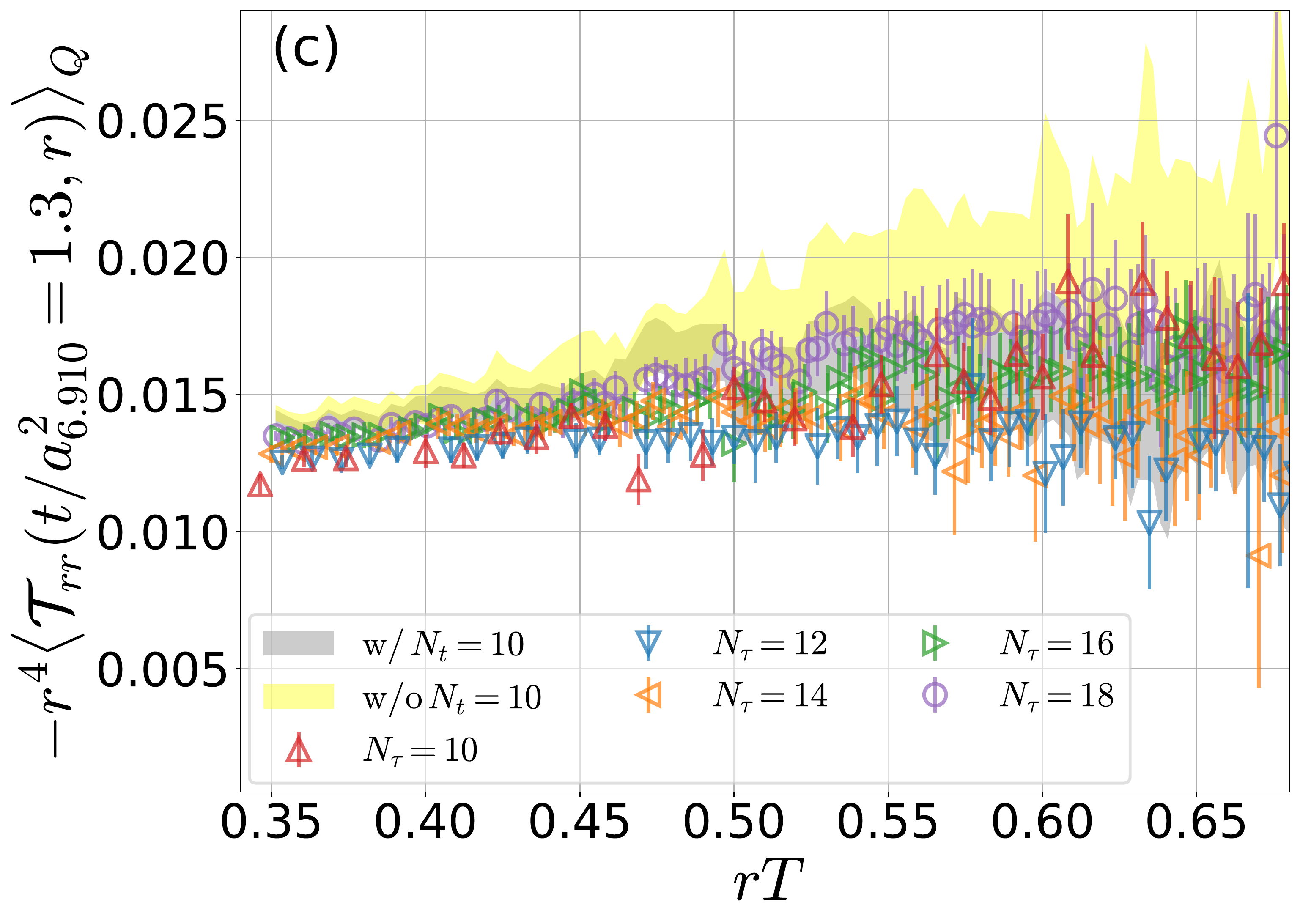} 
 \includegraphics[width=0.32\textwidth, clip]{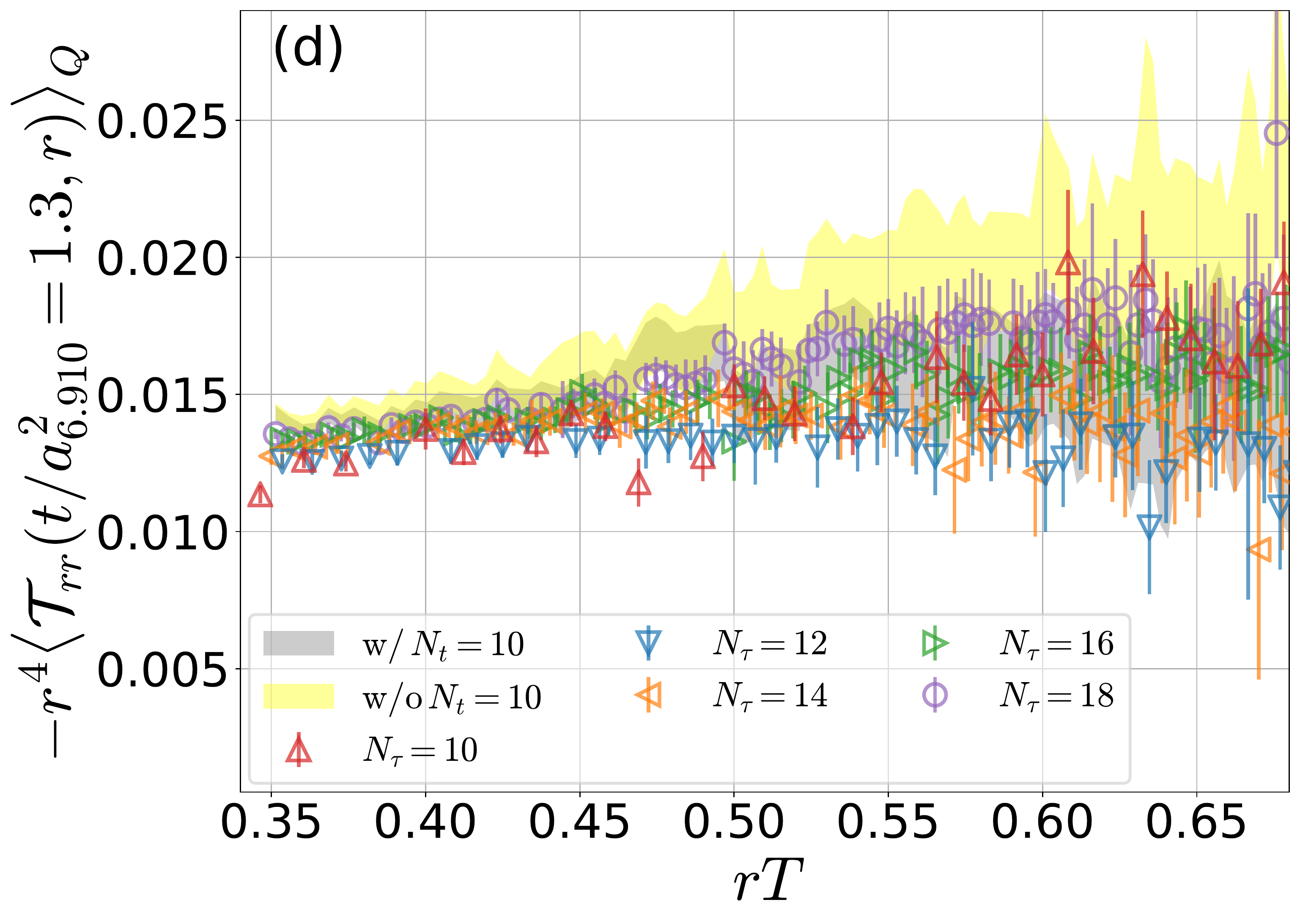}
 \caption{
 Distribution of 
 $-r^4\langle \mathcal{T}_{rr}(t/a_{6.910}^2=1.3,r) \rangle_Q$
 as functions of $rT$.
 (a) Same figure as Fig.~\ref{fig:finite_size_main}
 shown for a comparison.
 (b,c,d)
 Tree-level improved results with
 Eqs.~(\ref{eq:Eimp}) and (\ref{eq:Uimp_practice})
 for three choices of $U^\mathrm{tl}_{\gamma\gamma'}$:
 (b) $U^\mathrm{tl}_{\gamma\gamma'}=\langle U_{44}\rangle_Q$,
 (c) $U^\mathrm{tl}_{\gamma\gamma'}=\langle U_{rr}\rangle_Q$,
 (d) $U^\mathrm{tl}_{\gamma\gamma'}=\langle U_{\theta\theta}\rangle_Q$.
 }
 \label{fig:finite_size_append}
\end{figure}

As shown in Fig.~\ref{fig:finite_size_main},
there exists sizable discretization effect 
for the EMT distribution on coarse lattices especially for 
$N_{\tau}=10$.  
In this Appendix, we attempt to reduce such discretization effects
by using the tree-level lattice propagator.
Similar idea has been applied to the analysis of the Polyakov loop
correlations in Refs.~\cite{Necco:2001xg,Kaczmarek:2004gv}.

In calculating the EMT distribution around a static quark
Eq.~(\ref{eq:EMT-lat}),
we need
the expectation values of
Eqs.~(\ref{eq:E}) and (\ref{eq:U})
at nonzero flow time $t$.
These quantities are constructed from
$\langle G^a_{\mu\nu}(t,\bm{x}) G^a_{\rho\sigma}(t,\bm{x}) \rangle_Q$,
where the temporal coordinate is suppressed for notational simplicity.
In the continuum theory at the tree level, 
this operator is calculated to be
\begin{align}
 \langle G^a_{\mu\nu}(t,\bm{x}) G^a_{\rho\sigma}(t,\bm{x}) \rangle_Q
 = -g^2\frac{N^2-1}2 \mathcal{G}_{\mu\nu}(t,\bm{x}) \mathcal{G}_{\rho\sigma}(t,\bm{x}),
\end{align}
where $g$ is the gauge coupling and $N=3$ is the number of colors,
with
\begin{align}
  {\cal G}_{\mu\nu}(t,\bm{x}) \delta^{ab}
  = \int_0^{1/T} d\tau \langle A_4^a(\tau,\bm{0}) G_{\mu\nu}^b(t,x) \rangle.
\end{align}
By selecting appropriate gauge fixing conditions for the gauge action 
and the gradient flow equation, one obtains
$\mathcal{G}_{12}(t,\bm{x})=\mathcal{G}_{23}(t,\bm{x})=\mathcal{G}_{31}(t,\bm{x})=0$ and
\begin{align}
 & \mathcal{G}_{i4}(t,\bm{x})  = \partial_i D(t,\bm{x}), 
 \label{eq:<AF>}
 \\
 & D(t,\bm{x})
 = \int \frac{d^3 p}{(2\pi)^3} e^{i\bm{p}\cdot\bm{x}}
 \frac{e^{-tp^2}}{p^2}
 = \frac1{4\pi |\bm{x}|} {\rm erf}\Big(\frac{|\bm{x}|}{\sqrt{4t}}\Big).
 \label{eq:propagate_<AF>}
\end{align}

Next, in lattice gauge theory
the propagator corresponding to Eq.~(\ref{eq:propagate_<AF>})
with the Wilson gauge action for the gauge action and the flow equation
reads~\cite{Fodor:2014cpa,Altenkort:2020fgs}
\begin{align}
 {\cal D}(t, \bm{x}_{n})  
 = \int_{-\pi}^\pi \frac{d^3 p}{(2\pi)^3} e^{i \bm{p}\cdot\bm{x}_n}
 \frac{ e^{-t \sum_i \hat{p}_i^2 } }{ \sum_i \hat{p}_i^2 } ,
 \label{eq:hatFint}
\end{align}
with $\bm{x}_n = a\bm{n}= a(n_x,n_y,n_z)$ and 
$\hat{p}_i=(2/a) \sin (p_i/2a)$.
When the clover-leaf operator for the discretized representation
of $G_{\mu\nu}^a(t,\bm{x}_n)$ is employed, 
the discretized representation of 
$\mathcal{G}_{i4}(t,\bm{x}_n)$ is given by~\cite{Fritzsch:2013je}
\begin{align}
 \mathcal{G}^\mathrm{lat}_{i4}  (t,\bm{x}_n)  
 =  \frac{1}{2a} \left( {\cal D}(t, \bm{x}_{n+\hat{i}}) 
 - {\cal D}(t, \bm{x}_{n-\hat{i}}) \right).
 \label{eq:parallel}
\end{align}

Using Eq.~(\ref{eq:<AF>}) and (\ref{eq:parallel}), the tree-level
improvements of Eqs.~(\ref{eq:E}) and (\ref{eq:U}) denoted by
the superscript `imp' may be written as
\begin{align}  
& \langle E (t, \bm{x}_n) \rangle_Q^{\rm imp}
 = c(t,\bm{x}_n)    \langle E (t, \bm{x}_n) \rangle_Q, 
 \label{eq:Eimp} 
 \\ 
&  \langle {U}_{\gamma \gamma'}(t, \bm{x}_n) \rangle_Q^{\rm imp} =
 c(t,\bm{x}_n)  \langle {U}_{\gamma \gamma'}(t, \bm{x}_n) \rangle_Q ,
 \label{eq:Uimp}
\end{align}
where the correction factor $c(t,\bm{x}_n)$ is defined by
\begin{align}
 c(t, \bm{x}_n) = \frac{1}{3} \sum_{i=1}^3
 \Big( \frac{ \mathcal{G}_{i4} (t, \bm{x}_n) }{ \mathcal{G}_{i4}^{ {\rm lat}} (t,\bm{x}_n)}\Big)^2 .
 \label{eq:C}
\end{align}
In Eq.~(\ref{eq:C}), the average over $i$ is taken because generally
the ratio
$\mathcal{G}_{i4} (t, \bm{x}_n) / \mathcal{G}_{i4}^{ {\rm lat}} (t,\bm{x}_n)$
at a lattice site $\bm{x}_n$ depends on $i$.
However, in our particular choice of discretization,
i.e. the Wilson gauge actions and the clover-leaf operator,
it is easily shown that
$\mathcal{G}_{i4} (t, \bm{x}_n) / \mathcal{G}_{i4}^{ {\rm lat}} (t,\bm{x}_n)$
does not depend on $i$.
In this special case the average over $i$ in Eq.~(\ref{eq:C}) is redundant.
When this property is violated, the improvement of 
$\langle {U}_{\gamma \gamma'}(t, \bm{x}_n) \rangle_Q$ may be replaced by
the one that depends on $\gamma$ 
and $\gamma'$ in place of Eq.~(\ref{eq:Uimp}).

\begin{figure}[t]
 \centering
 \includegraphics[width=0.32\textwidth, clip]{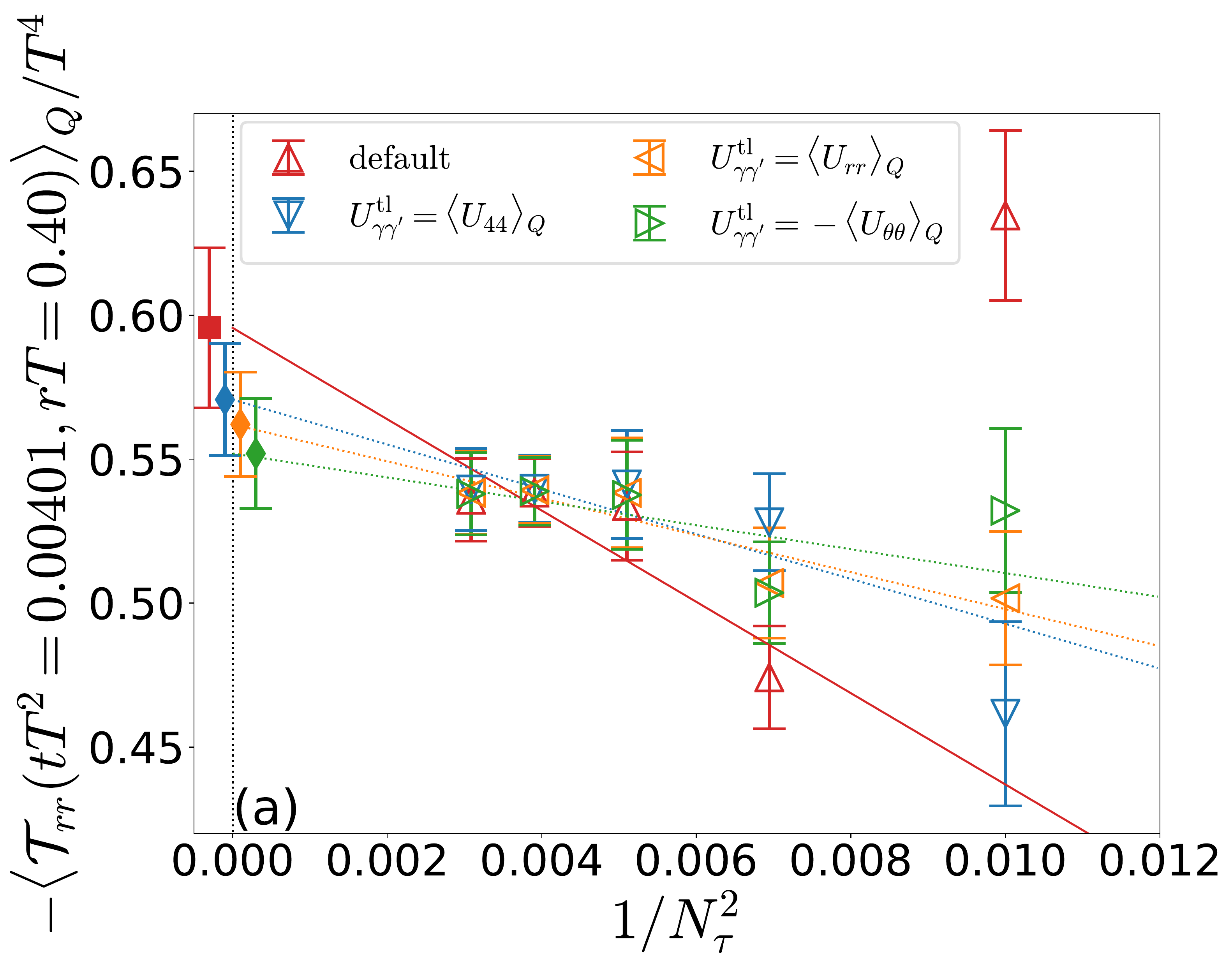}
 \includegraphics[width=0.32\textwidth, clip]{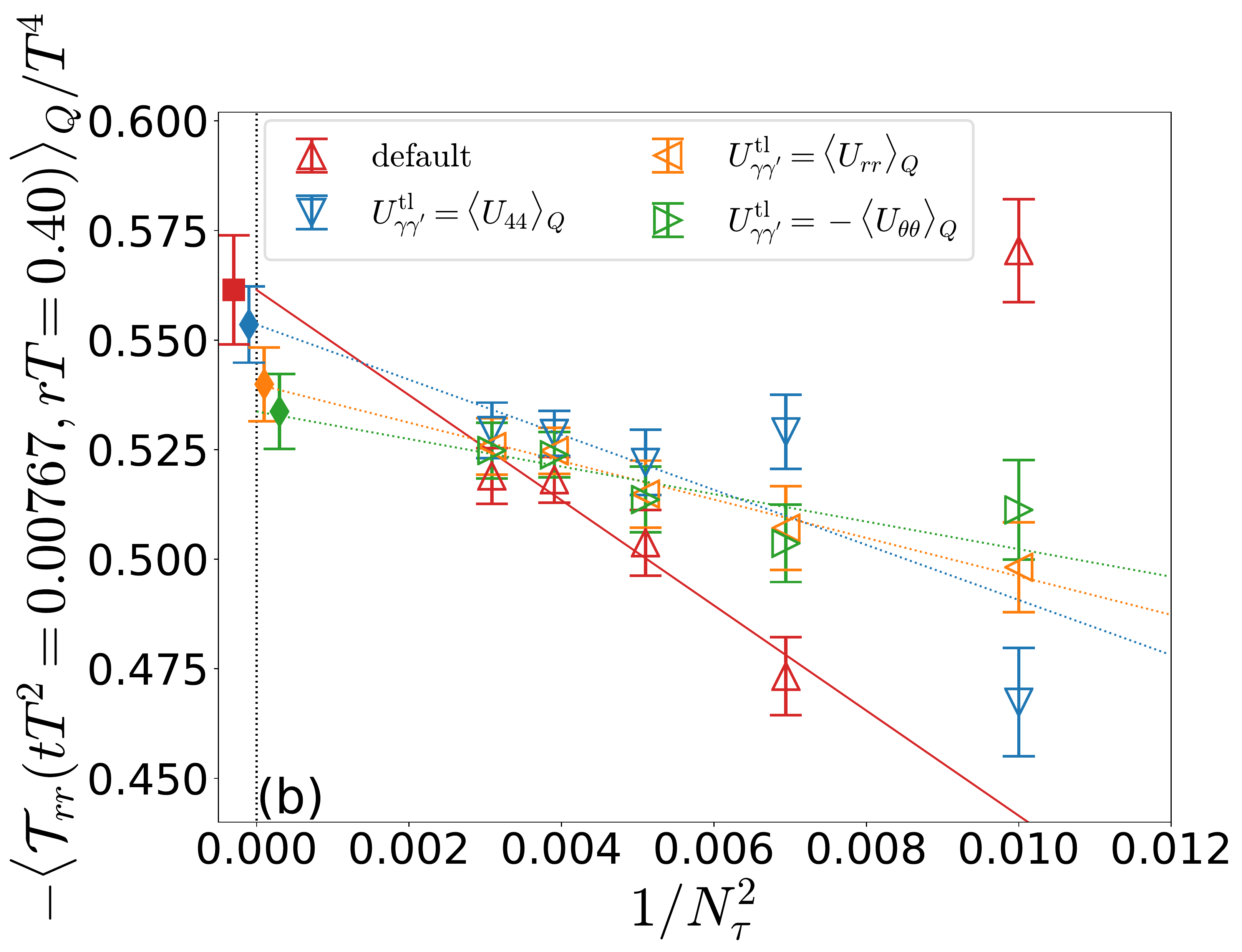}
 \includegraphics[width=0.32\textwidth, clip]{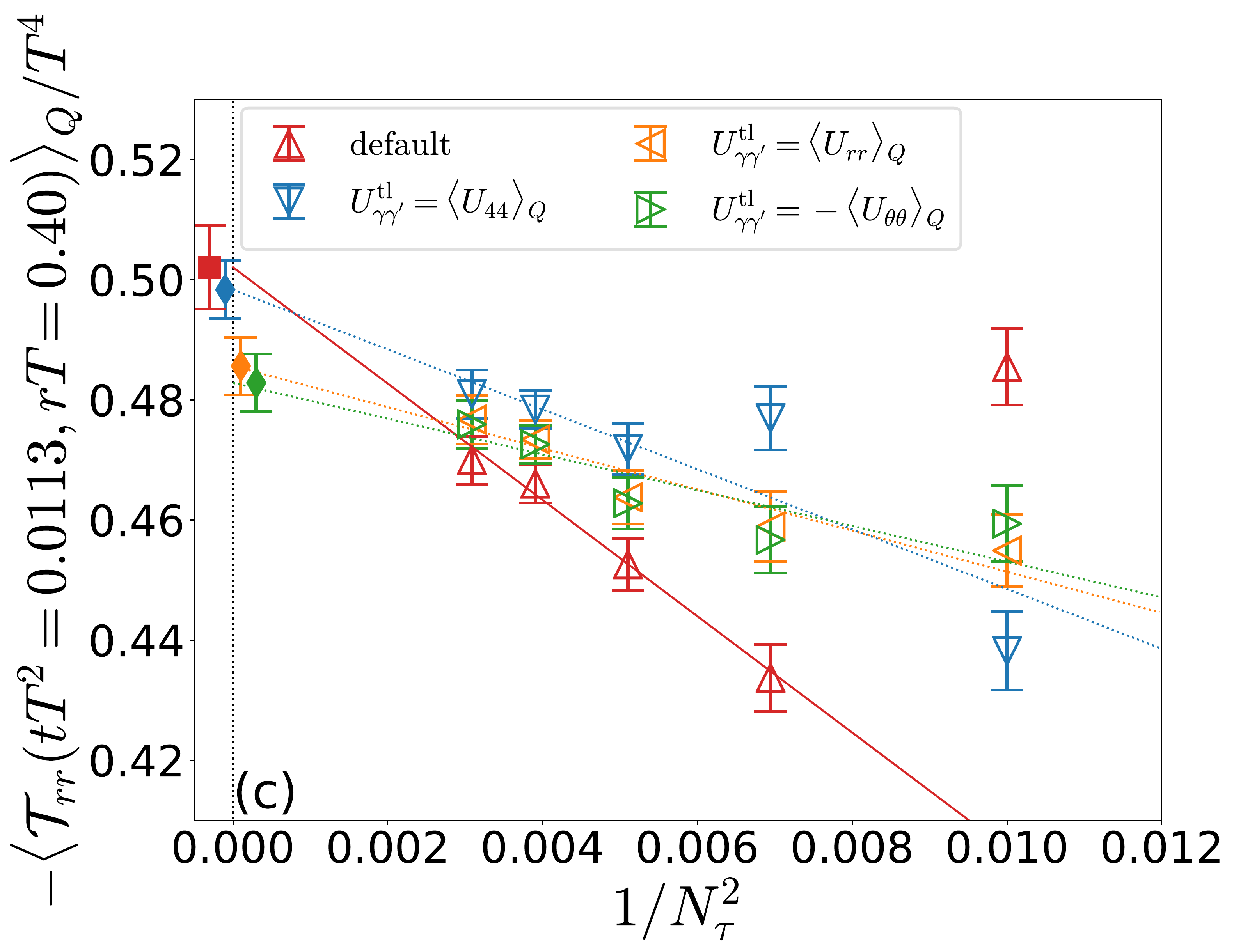}
 \includegraphics[width=0.32\textwidth, clip]{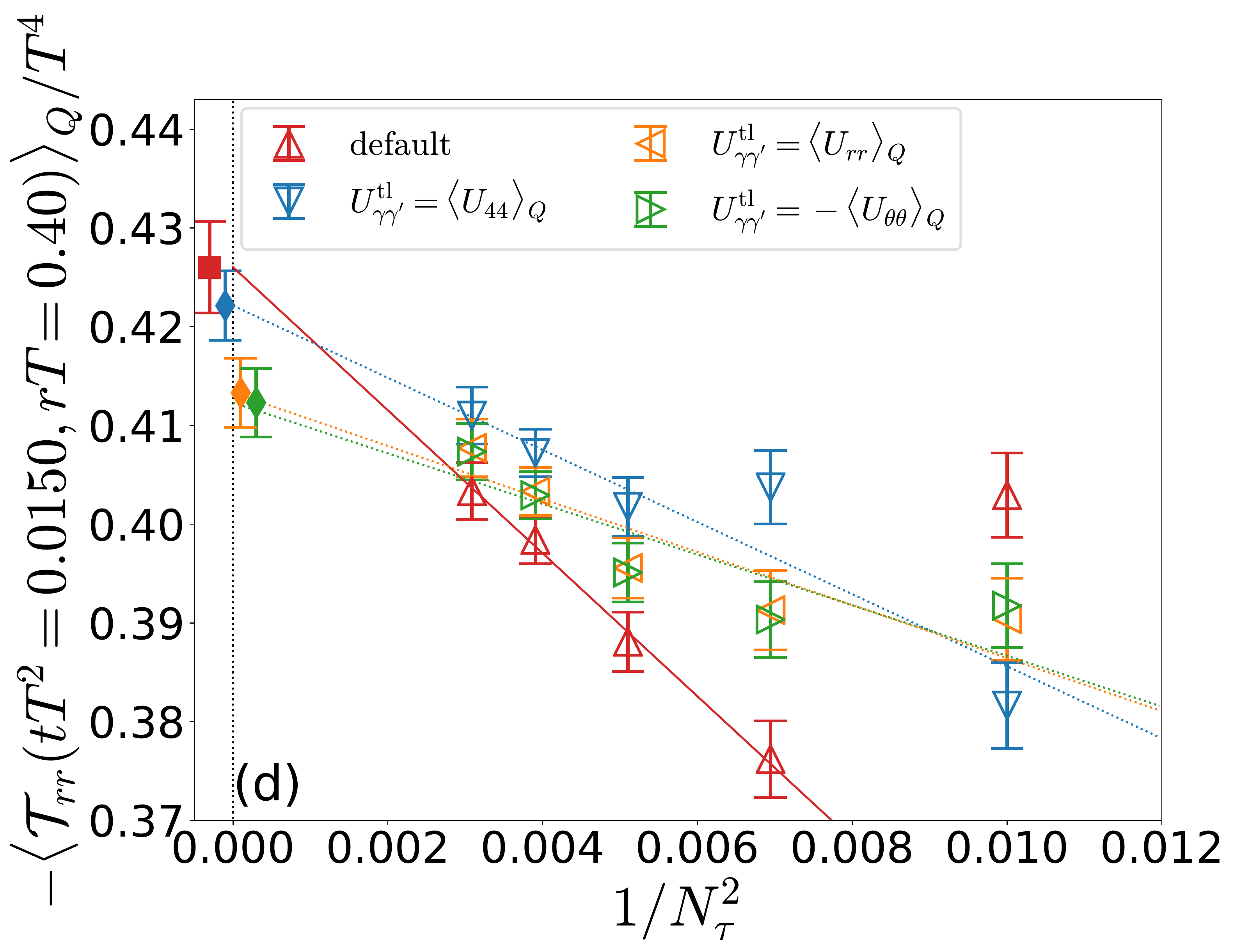}
 \caption{
 Comparison of the different prescriptions
 in the continuum extrapolation of 
 $-\langle {\cal T}_{rr} (t,rT=0.40)\rangle_{Q}/T^4$
 at $T/T_c=1.44$:
 (a) $tT^2=0.00401$,
 (b) $tT^2=0.00767$,
 (c) $tT^2=0.0113$, and 
 (d) $tT^2=0.0150$.
 }
 \label{fig:continuum_correction}
\end{figure}

There is one more subtle  issue about Eq.~(\ref{eq:Uimp}).
At the tree level, one easily finds that the matrix elements 
of $U_{44}$, $U_{rr}$ and $-U_{\theta\theta}$ 
are the same,  while the actual lattice data do not necessarily 
satisfy such relation as shown in the main text.
In our tree-level improvement, therefore, we decompose 
our lattice data into
tree-like part and the rest, 
$  \langle U_{\gamma \gamma'}(t, \bm{x}_n) \rangle_Q
   = U_{\gamma \gamma'}^{\rm tl}(t, \bm{x}_n)
   + \delta U_{\gamma \gamma'}(t, \bm{x}_n)$,
where the tree-like part $U_{\gamma \gamma'}^{\rm tl}(t, \bm{x}_n)$
satisfies $U_{44}^{\rm tl}(t, \bm{x}_n)=U^{\rm tl}_{rr}(t, \bm{x}_n)=-U_{\theta\theta}^{\rm tl}(t, \bm{x}_n)$.
Then we apply our  tree-level improvement only to the first term:
\begin{align}
 &  \langle {U}_{\gamma \gamma'}(t, \bm{x}_n) \rangle_Q^{\rm imp} =
 c(t,\bm{x}_n) U_{\gamma \gamma'}^{\rm tl}(t, \bm{x}_n)
 + \delta U_{\gamma \gamma'}(t, \bm{x}_n).
 \label{eq:Uimp_practice}
\end{align}  
Below, we consider three choices of 
$U_{\gamma \gamma'}^{\rm tl}(t, \bm{x}_n)$ to estimate 
the systematic uncertainty 
of this procedure: $U_{\gamma \gamma'}^{\rm tl}= \langle U_{44}
\rangle_Q, \langle U_{rr} \rangle_Q$,
and $- \langle U_{\theta\theta} \rangle_Q$.  
Corresponding results 
for $\langle {\cal T}_{rr}^\mathrm{R}(t,r) \rangle$ as example  
are shown in Fig.~\ref{fig:finite_size_append}(b), (c), and (d),
respectively,
together with the case without the correction,
Fig.~\ref{fig:finite_size_append}(a).
Colored open symbols represent the 
data at each $N_\tau$ and 
the gray (yellow) shade is the continuum result
with (without) the data at $N_\tau=10$.
The figures show that the tree-level improvement 
suppresses the discretization effect at short distances 
in three cases, especially for $N_{\tau}=10$.
    
Let us now compare  the continuum extrapolation 
using the data including $N_{\tau}=10$ 
with the tree-level improvement and that using 
the data without  $N_{\tau}=10$ and without 
the tree-level improvement. 
Shown in Fig.~\ref{fig:continuum_correction} is such a comparison for 
$-\langle {\cal T}_{rr} (t,rT=0.40)\rangle_{Q}/T^4$ 
at $T/T_c=1.44$
as functions of $1/N_\tau^2$.
Colored open triangles represent the data 
without the tree-level improvement (red) and 
with the tree-level improvement for three 
different prescriptions (blue, orange, and green).
Filled symbols at $1/N_\tau^2=0$ are the continuum extrapolation: 
The red squares are continuum results without $N_{\tau}=10$ data
discussed in the main text, while the diamonds are  
the continuum results with  $N_{\tau}=10$ data 
after the tree-level improvement.
Taking into the uncertainly associated 
with the different prescriptions for the tree-level improvement,
the default results without $N_{\tau}=10$  
in the main text are found to be consistent with the
improved results including $N_{\tau}=10$.

\section{Leading order perturbative analysis of EMT around a  static charge}
\label{sec:pert_calc}

Let us consider the SU($N$) Yang-Mills system 
at high temperature where $g(2\pi T) \ll 1$.
Then, the effective theory valid at the length scale 
of $R \gg (2 \pi T)^{-1}$  is 
the dimensionally reduced   electrostatic QCD (EQCD)  in  three dimensions
(see, e.g., \cite{Appelquist:1981vg,DHoker:1981bjo,Nadkarni:1982kb,Braaten:1994qx})
\begin{align}
 S_{\rm _{EQCD}} = \int d^3x\,\Bigl[\frac{1}{2}\mathrm{Tr}{\cal G}^2 
 + \mathrm{Tr}(D\varphi)^2
 + m_{\rm _D}^2\mathrm{Tr}\varphi^2 + \delta {\cal L}_{\rm _{EQCD}}\Bigr].
\end{align}
Here $({\cal A}_i, \varphi) = ({\cal A}^a_i t^a , \varphi^a t^a)=(A_i, A_4)/(g\sqrt{T})$, 
${\cal G}_{ij}= \partial_i{\cal A}_j - \partial_j{\cal A}_i + ig_{\rm _E}[{\cal A}_i, {\cal A}_j]$, 
and  $D_i\varphi = \partial_i\varphi + ig_{\rm _E} [{\cal A}_i, \varphi]$ 
with ${\rm Tr}(t^a t^b)=\frac{1}{2} \delta^{ab}$.
The higher dimensional operators are denoted 
by $\delta {\cal L}_{\rm _{EQCD}}$.
The  effective coupling and the Debye screening mass 
in the leading-order (LO)  
read $g_{\rm _E}= g\sqrt{T}$ and 
$m_{\rm _D}^2 = (N/3) (gT)^2$, respectively. 

\begin{figure}
 \centering
 \includegraphics[width=0.13\textwidth, clip]{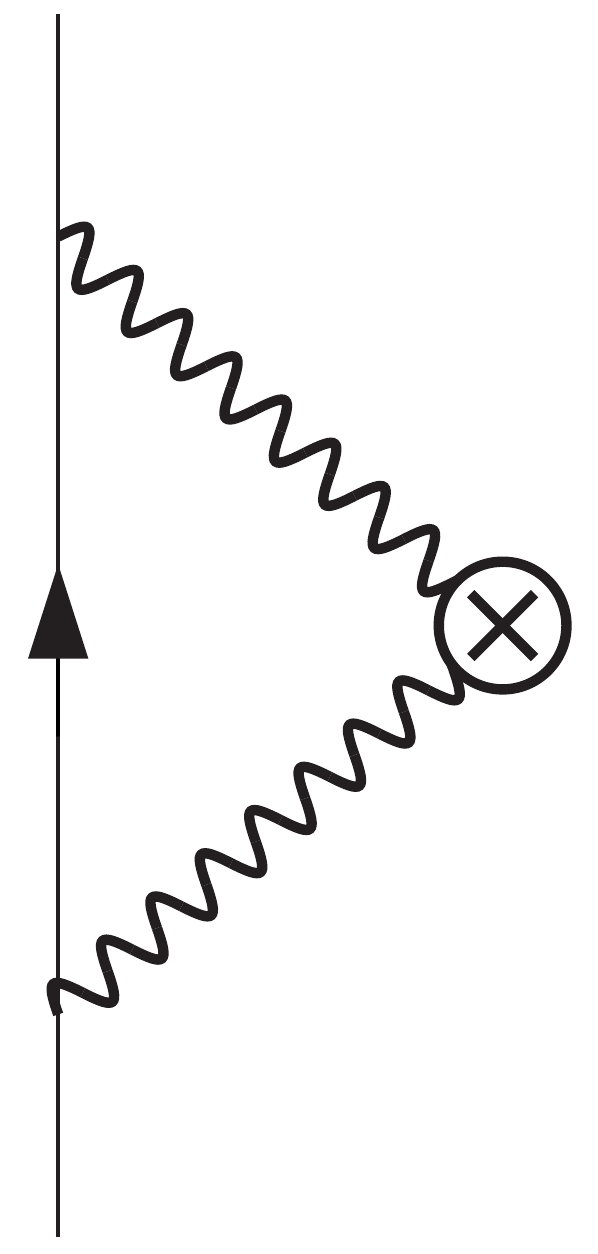}
 \caption{
 Diagram contributing to the leading-order calculation of the 
 correlation function between the Polyakov loop and 
 the EMT operator.
 The vertical line represents the Polyakov loop and two wavy lines 
 exchanged gluons. The symbol $\otimes$ corresponds to the EMT operator.
 }
 \label{fig:diagram}
\end{figure}

Under the ``Feynman static gauge'' ($\partial_4 {\cal A}_4=0$ for the 
temporal component and the 
Feynman gauge for the spatial  component 
${\cal A}_i$)  ~\cite{DHoker:1981bjo,Nadkarni:1982kb}, the 
tree-level propagators read
\begin{align}
 &\langle \varphi^a(\bm{0}) \varphi^b(\bm{x}) \rangle 
 = \delta_{ab} \frac{e^{-m_{\rm _D}|\bm{x}|}}{4 \pi |\bm{x}|},
 \\
 &\langle {\cal A}_i^a(\bm{0}){\cal A}_j^b(\bm{x}) \rangle 
 = \delta_{ab} \delta_{ij}  \frac{1}{4 \pi |\bm{x}|},
 \\
 &\langle {\cal A}_i^a(\bm{0}) \varphi^b(\bm{x}) \rangle = 0,
\end{align}
where  $a$ and $b$ are color indices.  
Moreover, the Polyakov loop operator  $\Omega(\bm{x})$ is written as 
\begin{align}
 \Omega(\bm{x})
 =\mathcal{P} e^{-ig\int_0^{1/T} d\tau A_4(\bm{x}, \tau)}
 = e^{-ig\varphi(\bm{x})/\sqrt{T}}.
\end{align}

The  leading order (LO)  contribution to the 
connected correlation between the Polyakov loop and the 
EMT   stems from the two gluon exchange of $O(g^2)$
and   is   diagrammatically
shown in  Fig.~\ref{fig:diagram}.  
Since the Polyakov loop operator has only the scalar component
$\varphi(\bm{x})$,  the terms which survive in the LO 
are the connected diagrams with $G_{4i}^2$, i.e., 
\begin{align}
 \langle (G_{i4}^a)^2 (\bm{x})  \rangle_Q
 = \frac{\langle (G_{i4}^a)^2 (\bm{x})   \mathrm{Tr}\Omega(\bm{0})\rangle_c}
 {\langle \mathrm{Tr}\Omega(\bm{0}) \rangle},
 \label{eq:cor_pol_emt}
\end{align}
where the suffix $c$ implies the connect correlation.

By expanding $(G_{i4}^a)^2$ and  $\mathrm{Tr}\Omega$  up to $O(\varphi^2)$
for fixed $i=1,2,3$, we obtain 
\begin{align}
 & \langle (G_{i4}^a)^2 (\bm{x})  \rangle_Q
 \notag
 \\
 &=
 -\frac{1}{4N}g^2 
 \langle \varphi^a (\bm{0}) \partial_i \varphi^b (\bm{x}) \rangle
 \langle \varphi^b (\bm{0}) \partial_i \varphi^a (\bm{x}) \rangle 
 + \mathcal{O}(g^3)
 \notag
 \\
 &=
 -\frac{N^2-1}{4N}g^2
 \left\{
 \partial_i\left(
 \frac{e^{-m_{\rm _D}|\bm{x}|}}{4\pi |\bm{x}|}
 \right)
 \right\}^2 + \mathcal{O}(g^3) \notag
 \\
 &=
 -\frac{C_F}{8\pi}\alpha_s
 \frac{x_i^2(m_{\rm _D} |\bm{x}|+1)^2}{|\bm{x}|^6}e^{-2 m_{\rm _D}|\bm{x}|} 
 + \mathcal{O}(g^3),
 \label{eq:ff_eqcd}
\end{align}
where $\alpha_s=g^2/4\pi$ and $C_F=(N^2-1)/2N$.

Picking up the contributions of $(G_{4i}^a)^2 (\bm{x})$ 
in  each component of the EMT, we
obtain the following perturbative estimate  
for $r \equiv | \bm{x}| \gg (2 \pi T)^{-1}$ up to $O(g^2)$,
\begin{align}
 &\langle
 \mathcal{T}_{44}(\bm{x}) \rangle_Q 
 =   \langle \mathcal{T}_{rr}(\bm{x}) \rangle_Q
 =  - \langle \mathcal{T}_{\theta \theta}(\bm{x}) \rangle_Q
 \notag
 \\
 & \ \ \ \ = -  \frac{C_F}{8\pi}\alpha_s
 \frac{(m_{\rm _D} r +1)^2}{r^4}e^{-2 m_{\rm _D}r} + \mathcal{O}(g^3),
 \label{eq:each_eqcd}
\end{align}
Simplest way to show the above 
relation is to choose $\bm{x} = (r, 0, 0) $, so that
$\mathcal{T}_{rr}(r,0,0) = \mathcal{T}_{11}(r,0,0)$ and
$\mathcal{T}_{\theta \theta}(r,0,0) = \mathcal{T}_{22}(r,0,0)$.
  
Although one finds that the EMT trace  
$  \langle \mathcal{T}_{\mu\mu}(\bm{x})  \rangle_Q$ vanishes at $O(g^2)$,
one can utilize the following trace anomaly to evaluate 
its $O(g^4)$ contribution:
\begin{align}
 \mathcal{T}_{\mu\mu} = \frac{\beta}{2g}G_{\mu\nu}^aG_{\mu\nu}^a,
\end{align}
where the Yang-Mills beta function reads 
$\beta = -\beta_0 g^3 - \beta_1 g^5 + \cdots$,
with  $\beta_0 = (11/3) C_A /(4\pi)^2$,
$\beta_1 = (34/3)C_A^2/(4\pi)^4$, and 
$C_A=N$.  By using the right hand side of the formula and follow the same 
procedure as above, we find 
\begin{align}
 \langle \mathcal{T}_{\mu\mu}(\bm{x})  \rangle_Q
 =
 - \frac{11}{3} \frac{C_A C_F}{(4\pi)^2}\alpha_s^2  
 \frac{(m_{\rm _D} r +1)^2}{r^4}e^{-2 m_{\rm _D}r} + \mathcal{O}(g^5). 
 \label{eq:tr_eqcd}
\end{align}
This is indeed $O(g^2)$ higher  than Eq.~(\ref{eq:each_eqcd}).

\end{document}